\documentstyle[epsf,epsfig,rotate]{aa}

\newcommand{\be}{\begin{equation}}
\newcommand{\ee}{\end{equation}}

\newcommand{\reference}{\bibitem}

\newcounter{pp3}
\addtocounter{pp3}{3}

\begin{document}

\title
{Models of Accreting Gas Giant Protoplanets
in Protostellar Disks}

\author{John C.B. Papaloizou \& Richard P. Nelson}
\institute{ Astronomy Unit,
Queen Mary, University of London, Mile End
Rd, London, E1 4NS, U.K.}

\offprints{jcbp@maths.qmw.ac.uk}

\date{Received /Accepted}

\def\LaTeX{L\kern-.36em\raise.3ex\hbox{a}\kern-.15em
T\kern-.1667em\lower.7ex\hbox{E}\kern-.125emX}
\titlerunning{Protoplanet models}
\authorrunning{J.C.B. Papaloizou \& R.P. Nelson }

\abstract{
We present evolutionary models of gas giant planets forming in
protoplanetary disks. We first consider protoplanet models
that consist of solid cores surrounded by hydrostatically supported
gaseous envelopes that are in contact with the boundaries  of 
their Hill spheres, and
accrete gas from the surrounding disk. We neglect planetesimal accretion, and
suppose that the luminosity arises from gas accretion alone.
This generally occurs on a long time scale which may be
comparable to the protostellar disk lifetime.
We classify
these models as being of type A, and follow their quasi static evolution
until the point of rapid gas accretion is reached.

We consider a second class of protoplanet models that have not hitherto 
been considered. These models have a free surface, their energy supply
is determined by gravitational contraction,  and  mass  accretion
from the protostellar  disk 
that is assumed to  pass through a circumplanetary disk.
An evolutionary sequence is obtained by specifying the accretion rate
that the protostellar  disk is able to supply.
We refer
to these models as being of type B. An important
result is that these protoplanet models
contract quickly  to a radius $\sim 2\times 10^{10}cm$
and are able to accrete
gas from the disk at any
reasonable rate that may be supplied without any consequent  expansion
(e.g. a Jupiter mass in $\sim$ few $\times 10^3$ years, or more slowly
if so constrained by the disk model).
We speculate that the early stages of gas giant 
planet formation proceed along evolutionary paths described by models
of type A, but at the onset of rapid gas accretion 
the protoplanet contracts interior to its Hill sphere,
  making a transition to an evolutionary path
described by models of type B, receiving gas  through a circumplanetary
disk that forms within its Hill sphere, which is in turn
fed by the surrounding protostellar disk.

We consider planet models with solid core masses of 5 and 15 M$_{\oplus}$,
and   consider evolutionary sequences
assuming different amounts of dust opacity
 in the gaseous envelope.
The   initial protoplanet  mass doubling time scale is very approximately
 inversely proportional
to the magnitude of  this  opacity.
Protoplanets with 5 M$_{\oplus}$ cores, and standard dust  opacity 
require $\sim 3 \times 10^8$  years to grow to a Jupiter mass,
longer than reasonable disk life-times. 
A model with  1 \% of standard dust opacity requires 
 $ \sim 3\times 10^6$ years.
Rapid gas accretion in both these cases ensues once the planet
mass exceeds $\simeq 18$ M$_{\oplus},$ with substantial time
spent in that mass range. 

\noindent Protoplanets with 15 M$_{\oplus}$ cores grow to a Jupiter mass
in $\sim 3 \times 10^6$ years if standard dust  opacity is assumed, 
and in $\sim   10^5$
years if 1 \% of  standard dust opacity is adopted.
In these  cases , the planet
spends substantial time with mass between 30 -- 40 M$_{\oplus}$ before
making the transition to rapid gas accretion.  We  emphasize that these growth times
apply to the gas accretion phase and not to  the prior  core formation phase.

\noindent   According to  the usual theory of protoplanet migration,
although there is some dependence on disk parameters,
 migration in standard model disks 
is most effective in the mass range where the transition
from type A to type B occurs. This is also
 the transitional regime
between  type I  and type II migration. If a mechanism 
prevents the type I migration of low mass protoplanets, they
could then undergo a rapid inward migration at around the transitional
mass regime. Such protoplanets would end up in the inner regions of the
disk undergoing type II migration and further accretion potentially
becoming sub Jovian close orbiting planets. Noting that more dusty
and higher mass cores spend more time at a larger transitional 
mass that in general  favours more rapid migration,  such planets are more likely
to become close orbiters.

\noindent  We find that the luminosity of the forming protoplanets
 during the later stages of gas accretion is dominated by the circumplanetary
disk and protoplanet-disk boundary layer.
 For final accretion times for one Jupiter mass
in the range $10^{5-6}y,$ the luminosities are in the range
 $\sim 10^{-(3-4)}  L_{\odot}$ and the characteristic 
temperatures are in the range $1000-2000K.$
 However, the luminosity may reach $\sim 10^{-1.5}  L_{\odot}$
for shorter time periods at the faster rates of accretion
that could be delivered by the  protoplanetary disk.

\vspace*{0.5cm}
\keywords{accretion, accretion disks --- solar system: formation ---
planetary systems }}

\maketitle

\section{Introduction}
\label{sec:intro}
Planets are believed to form out of protostellar disks by either
gravitational instability (Cameron 1978; Boss 1998) or by
a process of growth through planetesimal accumulation followed, in the
giant planet case, by gas accretion (Safronov 1969; Wetherill~\&
Stewart 1989; Mizuno 1980). It is the latter mechanism that 
we consider in this paper.

\noindent  The process is presumed to begin
with the accumulation of the solid cores by the
accretion of planetesimals typically exceeding
a kilometer in radius  which have been formed  
through the collisional growth and sedimentation of dust grains in
the protoplanetary disk (see Lissauer 1993 and references therein).
Once the solid core becomes massive enough 
a significant gaseous atmosphere forms. 
The mass required depends to some extent on physical conditions
in the disk, the rate of planetesimal
accretion and the dust opacity but is typically several earth
masses 
(eg. Mizuno 1980; Stevenson 1982; Bodenheimer \& Pollack 1986).
This is consistent with models of
Jupiter which indicate that it has a solid core typically  of
this magnitude (Podolak et al. 1993). We note, however, that more recent models
suggest that Jupiter's core may be less massive than previously thought
(Saumon \& Guillot 2004). Models of Saturn still indicate a core mass of
$\simeq 10$ M$_{\oplus}$.

\noindent During the early build up of the core the luminosity
is due to the liberation of gravitational energy by accreting planetesimals.
However, once the 
mass of the gaseous envelope starts to become significant 
the gravitational settling of the gas becomes important and 
at some cross over point becomes dominant
(Pollack et al. 1996). At this point models assuming strict thermal
equilibrium break down. This is manifest through the fact that
for fixed luminosity due to planetesimal accretion, there is a maximum
or critical core mass for which a strict thermal equilibrium
model can be constructed (see eg. Papaloizou \& Terquem 1999).
Beyond this point  the evolution is no longer in thermal
equilibrium and if the protoplanet remains in contact
with adequate disk material, gas accretion may ensue.

\noindent The purpose of this paper is to examine
the protoplanet models subsequent to the attainment
of the critical core mass in the context
of the protoplanetary disk environment
and disk planet interactions. We assume that the core becomes isolated
from further planetesimal accretion so that settling of accreted gas
is the only energy source. The rationale for this assumption is discussed
in section~\ref{sec:typeB-evol}. We consider two types of model.
The first type, which we denote as type A, is fully embedded 
in the protostellar disk
and hence has an effective radius equal to that of the Roche lobe
or Hill sphere. This is the correct radius to use rather than the 
Bondi radius which is never significantly smaller for any of the models
we study. At some mass, these models tend to 
enter a rapid accretion phase. This occurs when the planet mass $\sim 0.1M_J,$
$ M_J$ denoting a Jupiter mass, and 
is similar to that for which either  significant perturbation
to the protoplanetary disk through local mass 
accretion or disk--planet interaction begins (e.g. Nelson et al. 2000).
These processes eventually lead to gap formation.
Accordingly we consider models of a second type , type B,
which are no longer enveloped at the Roche lobe but are assumed to
have a free surface and accrete from a circumplanetary disk
at a rate determined by the external throughput from the protostellar disk. 
We find that these can be constructed for a wide range of accretion rates
indicating that during the later stages a forming protoplanet
can  comfortably 
accrete at any rate reasonably supplied by the protostellar disk.

We supplement these models of protoplanet evolution with hydrodynamical 
simulations of the interaction between low mass protoplanets and
protostellar disks. The purpose of these models is to establish
plausible accretion times scales for the freely accreting protoplanet
models of type B.

This paper is organised as follows.
We present the basic equations for the protoplanet models in 
section~\ref{sec:envelope_eq}, and discuss the appropriate
boundary conditions in section~\ref{sec:bcs}.
In section~\ref{sec:Accretion} we describe how evolutionary
sequences are constructed for protoplanet models
of type A and B, accounting for gas accretion from
the protostellar disk. We discuss the numerical procedure adopted
for the hydrodynamic simulations of disk-planet interactions in
section~\ref{sec:disk-planet}. The results of our calculations
are presented in section~\ref{sec:envelope_calc}, and their implications
are discussed in section~\ref{sec:discussion}. Finally we draw our conclusions
in section~\ref{sec:conclusion}.

\section{Basic Equations for the Protoplanet Models}
\label{sec:envelope_eq}
We adopt the approach of previous workers (eg. Bodenheimer \& Pollack 1986;
Pollack et al. 1996) and approximate the protoplanetary structure as
being spherically symmetric, the approach  being similar to that
followed in modeling stellar structure. Many of the details are given
in Papaloizou \& Terquem (1999). 

\noindent The interior state variables at any point in a  
model are functions only  of the distance to the centre,
$r,$  also characterized as the spherical polar radius. 
We assume the models are in 
hydrostatic  equilibrium and neglect rotation.  The
equation of hydrostatic equilibrium is 

\begin{equation}
\frac{dP}{d r} = - \rho g .
\label{dpdvarpi}
\end{equation}

\noindent Here, the pressure is $P,$
the  local acceleration due to gravity
is $g=G M(r) / r^2$ with
$M(r)$ being the mass, including that of any solid core,  interior to radius $r$
and $G$ is the gravitational constant.
The mass interior to radius $r$ satisfies

\begin{equation}
\frac{dM}{d r} = 4 \pi r^2 \rho,
\label{dmdvarpi}
\end{equation}
where $\rho$ is the density.

\begin{figure*}
\epsfig{file=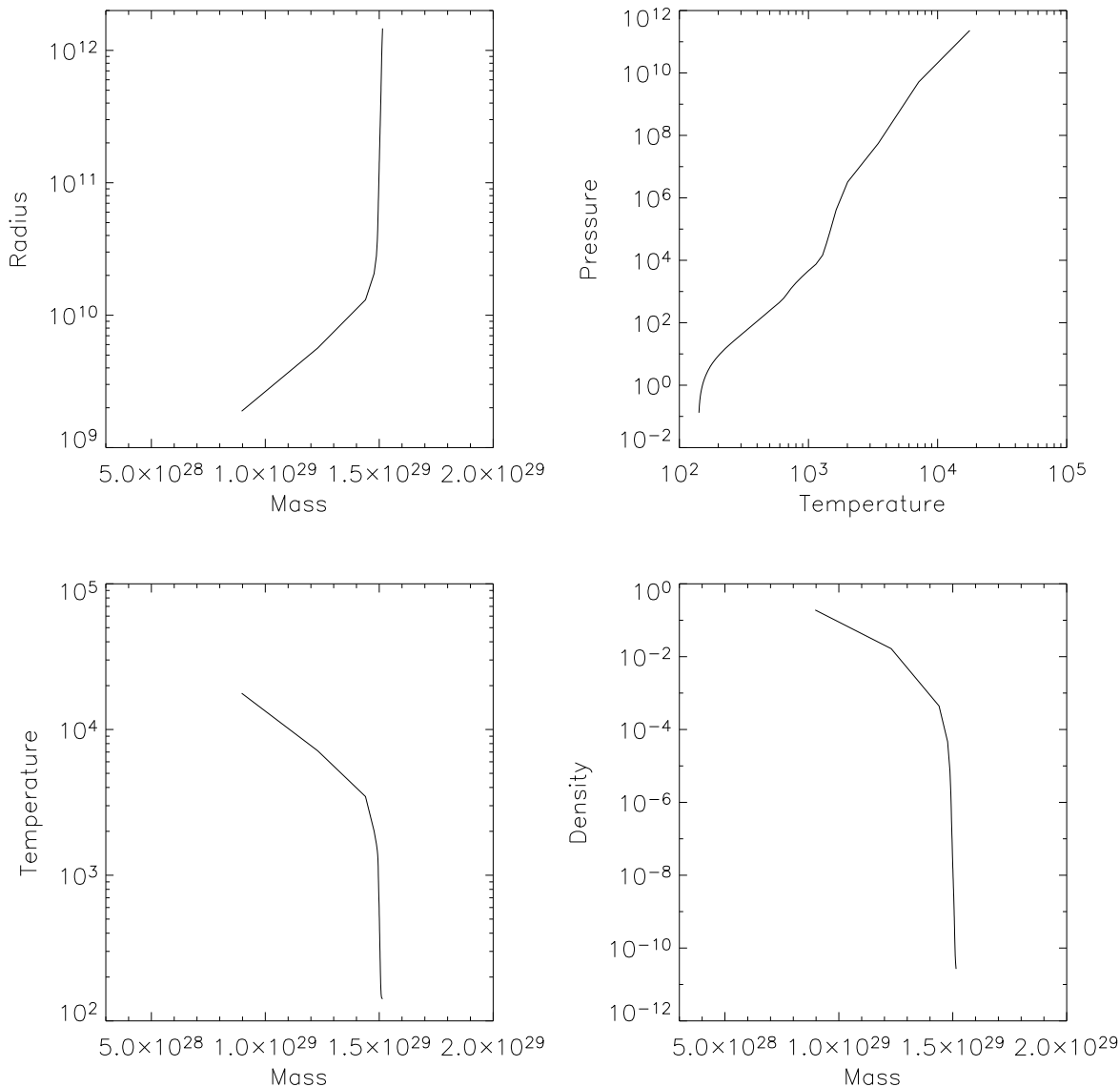, height=16cm, width=16cm,angle=360}
\caption[]{ State variables are plotted for a protoplanet model
of total mass  $ 25.3 M_{\oplus}$ with a $15M_{\oplus}$ solid
core.  Apart from
temperature in K,
these are given in cgs units.
The upper left panel gives a plot of interior radius
as a function of interior mass.
The upper right panel gives a pressure temperature plot.
The lower left panel gives a plot of
 temperature
as a function of interior mass.
The lower right panel gives a plot of
the local density
  as a function of interior mass.
Convective  heat
transport occurs when
$679K > T  > 263K $ and when $T > 2100K. $
 Approximately the gas component of the
inner  ninety  eight percent of the mass
is convective.}
\label{fig0}
\end{figure*}

\begin{figure*}
\epsfig{file = 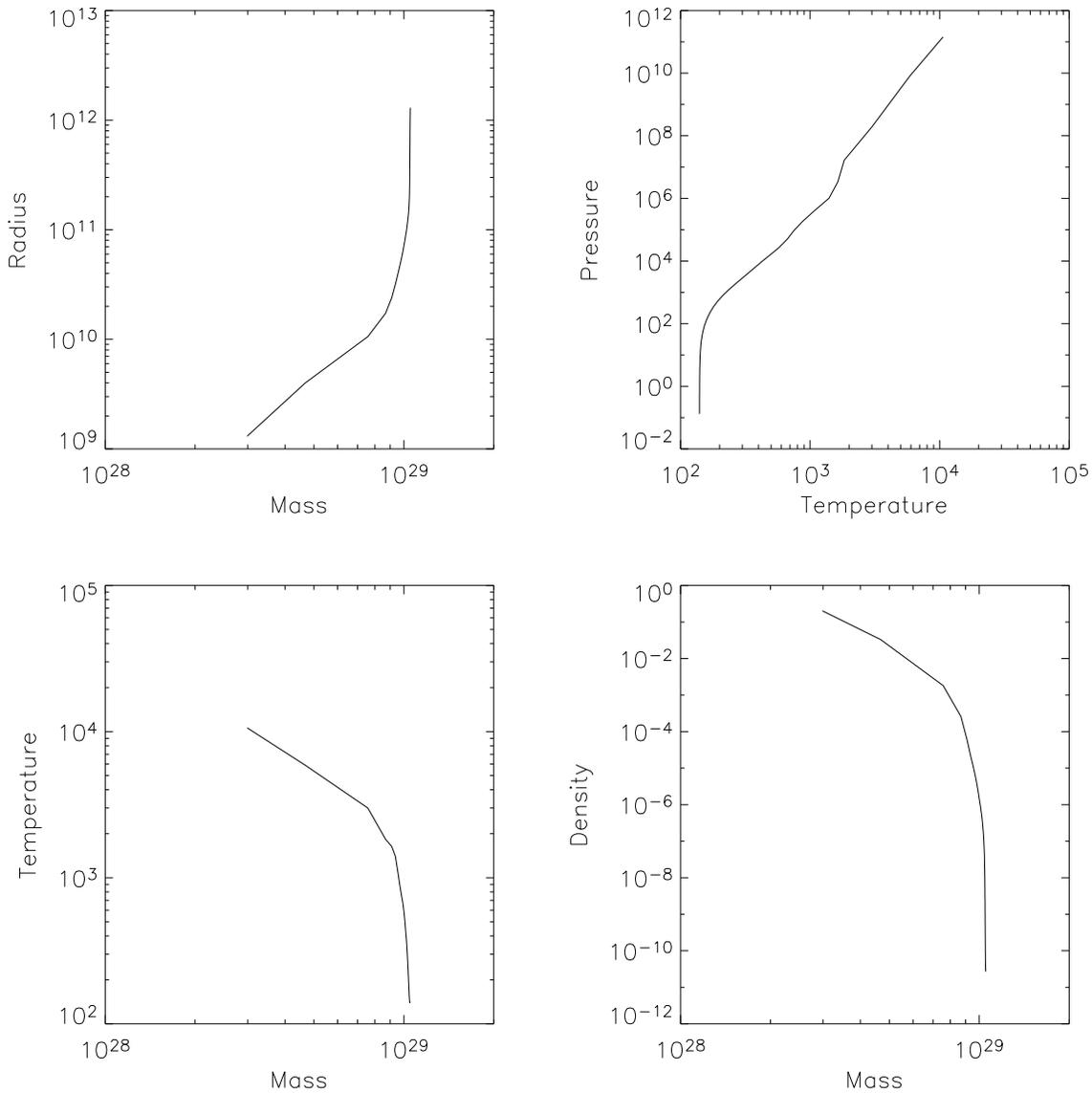, height=16cm,  width=16cm, angle=360}
\caption[]{ As in figure \ref{fig0}, but
 for a protoplanet model
of total mass $ 17.6 M_{\oplus}$ with a $5M_{\oplus}$ solid core.
Convective  heat
transport occurs when
$720K > T  > 264K $ and when $T > 2100K. $
 Approximately the gas component of the  the  inner  eighty percent of the mass
is convective.
}
\label{fig1}
\end{figure*}

\begin{figure*}
\epsfig{ file = 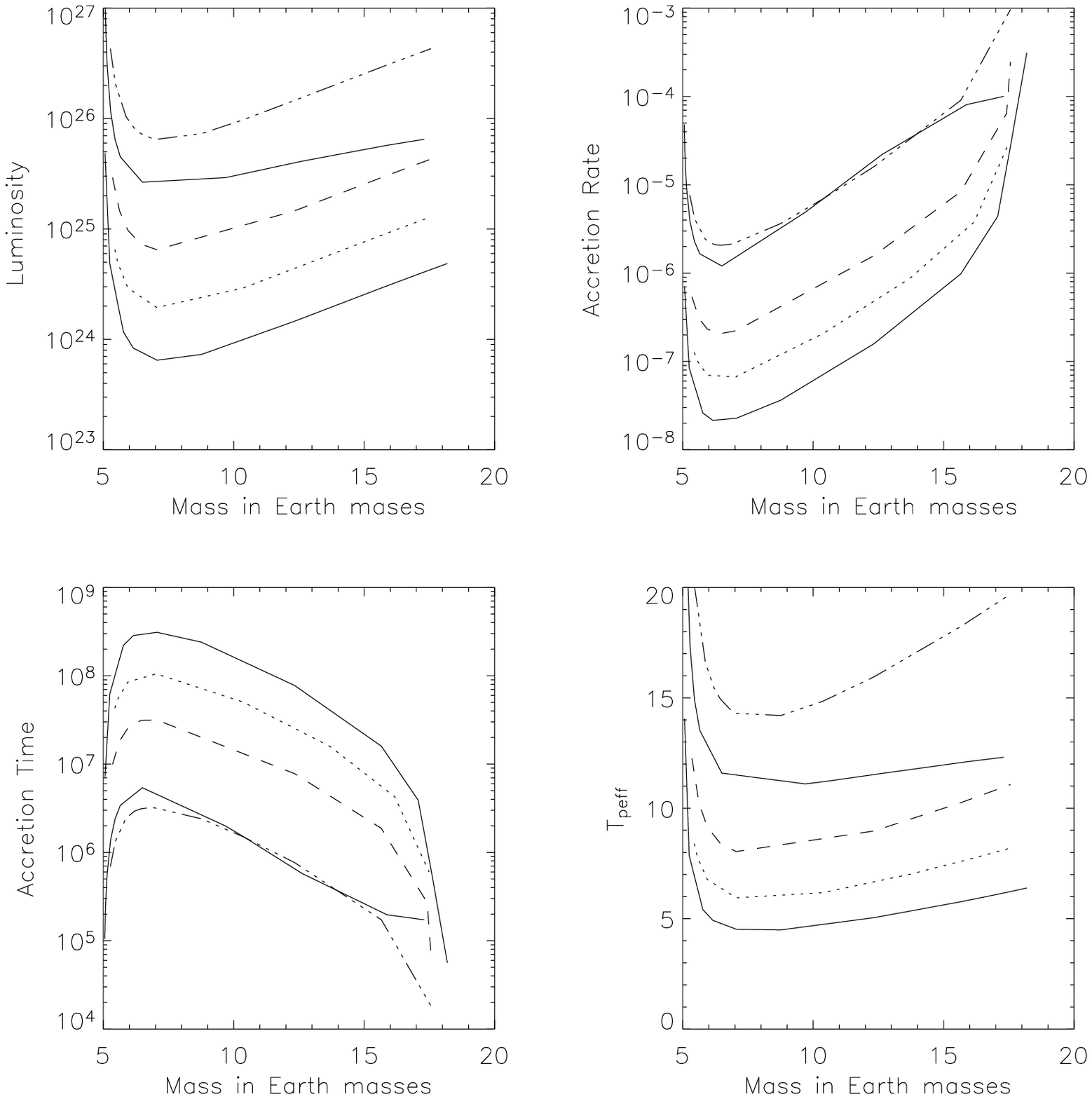, height=16cm,  width=16cm, angle=360}
\caption[]{ This figure illustrates the  evolution of
protoplanet models which maintain contact with the
protoplanetary disk and   fill their Roche lobes while
they accrete from it. They have fixed
solid core masses  of $5 M_{\oplus} $ and are situated at $5AU.$
The upper left panel shows Luminosity in cgs units
 as a function of their increasing mass, $M,$ in  earth masses.
The upper right panel shows the gas accretion rate in
$M_{\oplus} y^{-1}$  as a function of mass
while the lower left panel gives the accretion time $ M / {\dot M}$ in $yr$
as a function of mass. The lower
right panel gives the temperature $T_{effp}$ (see text) as a function of mass.
The   models shown have standard opacities (full line), standard opacities
reduced by a factor of three (dotted line), standard opacities
reduced by a factor of ten ( dashed line), and standard opacities reduced
by a factor of  one hundred ( triple dot dashed line)  respectively.
In addition models with an opacity reduction of a factor
of one hundred applied only for $T<1600K$ are plotted ( upper full line
in all panels except lower left where it is the lower full line.) }
\label{fig2}
\end{figure*}

\noindent For the calculations presented here, we adopt the
equation of state for a hydrogen and helium mixture given by Chabrier
et al. (1992).  The mass fractions of hydrogen and helium are taken to
be 0.7 and 0.28, respectively.  The luminosity $L_{rad}$ transported 
by radiation satisfies 

\begin{equation}
\frac{L_{rad}}{4 \pi r^2} = 
-\frac {4 ac
T^3} {3 \kappa \rho} 
\frac{dT}{d r}, 
\label{dtdvarpi}
\end{equation}
where $a,$ $c,$ $ T $ and $\kappa$ are the radiation constant,
the speed of light, the temperature
and the opacity respectively. The calculations
reported here are based on the opacities given by
Bell \& Lin (1994). These, being functions
of density and temperature,  include contributions from molecules, atoms, ions
and dust grains. The latter produce an increase in the opacity
amounting to several orders of magnitude for $T < 1600K.$

\subsection{Inner Convective Regions} 
Most of the gas mass within the models is unstable to convection
so a theory of energy transport by convection is needed.
We adopt the conventional mixing length theory (eg. Cox~\& Giuli 1968).

\noindent  
The radiative and adiabatic temperature gradients  $\nabla_{rad}$
and $\nabla_{ad}$ are defined through

\begin{equation}
\nabla_{rad} = \left( \frac{\partial \ln T}{\partial \ln P}
\right)_{rad} = \frac{3 \kappa L P}{16 \pi
ac G M T^4} ,
\label{dTdr_rad}
\end{equation}

\noindent and

\begin{equation}
\nabla_{ad} = \left( \frac{\partial \ln T}{\partial \ln P} \right)_S ,
\end{equation}

\noindent with the subscript $S$ denoting evaluation at constant
entropy. $\nabla_{ad}$ is a quantity determined directly from the thermodynamics
of the equation of state alone.

\noindent The total luminosity
is $L.$ During the phase of solid core growth
it is expected that this is produced by the   
gravitational energy of accreting planetesimals 
(e.g., Mizuno
1980; Bodenheimer \& Pollack 1986). However, for the later phases considered
here, the source of energy is primarily settling and accretion of gas
(see section~\ref{sec:typeB-evol}).

\noindent  When $\nabla_{rad} < \nabla_{ad}$, the gas is 
convectively stable and the energy is transported
entirely by radiation. On the other hand
when $\nabla_{rad} > \nabla_{ad},$ the medium is
convectively  unstable  and  some of the energy is transported by
convection.  We write the total luminosity
passing through a sphere of radius $r$ 
as    $L_r(r) = L_{rad}+L_{conv}$, where $L_{conv}$ is the
luminosity associated with convection. Adopting the mixing length theory 
 (Cox~\& Giuli 1968) we have 

\begin{eqnarray}
L_{conv} & = & \pi r^2 C_p \Lambda^2 \left[ \left( \frac{\partial
T}{\partial r} \right)_S - \left( \frac{\partial T}{\partial r}
\right) \right]^{3/2}
\nonumber \\
& \times & 
\sqrt{ \frac{1}{2} \rho g \left| \left(
\frac{\partial \rho}{\partial T} \right)_P \right| } , \label{mixl}
\end{eqnarray}

\noindent where $\Lambda=|\alpha P/(dP/d r)|$ is the mixing
length, $\alpha$ being a constant parameter  expected to be  of order unity, $\left(
\partial T/\partial r \right)_S = \nabla_{ad} T \left( d \ln P / d r
\right)$, and the subscript $P$  denotes
evaluation  at  constant pressure.  
 We adopt the mixing length parameter $\alpha  = 1$.

\section{Boundary Conditions}
\label{sec:bcs}
\subsection{ The Inner Boundary}
We  assume that there is a  solid core of mass $M_{core}$
with a uniform mass density
$\rho_{core}=3.2$~g~cm$^{-3}$ (eg. Papaloizou~\&   Terquem 1999). 
The  boundary  condition the models that calculate the structure
of the gaseous envelope must satisfy is
that the total mass  $ M(r_{core}) = M_{core}$ when
\begin{equation}
r= r_{core} = \left( \frac{3 M_{core}}{4 \pi \rho_{core}} \right)^{1/3}.
\end{equation}

\noindent For models with no source 
of accretion energy at the core surface, such as those considered here,
we also require $L_r=0.$
However, the model interiors are convective with radiation making
negligible contribution to the heat transport, accordingly adiabatic stratification
is a good approximation near the inner boundary independently
of any reasonable value for $L_r,$  with the consequence that we do not actually  need
to enforce the condition $L_r=0$ there.

\subsection{ The Outer Boundary}
We here consider two different classes of model which have differing boundary conditions.
We consider each of these in turn.

\subsubsection{Enveloped Models in Contact with the Roche Lobe}
For these models, subsequently denoted as of type A, we assume the structure extends to 
 the Roche lobe or boundary of the Hill sphere beyond which
material must be gravitationally unbound from the protoplanet.
 For this radius we adopt

\begin{equation}
r_L = \frac{2}{3} \left( \frac{M_{pl}}{3 M_{\ast}} \right)^{1/3} R_p ,
\end{equation}

\noindent where $M_{pl}$ is the total  planet mass 
including gas and solid core and 
 $R_p$ is the orbital
radius or distance  of the protoplanet, assumed in circular orbit,
from the central star. The structure state variables are assumed to eventually join
smoothly to those associated with the enveloping protoplanetary disk
where 
$ T= T_d,$ $ P= P_d$ and $ \rho = \rho_d$, respectively.

\noindent Thus the boundary conditions 
are that  at $r=r_L,$  $ M(r_L) = M_{pl}$, 
 $P = P_d$ and the temperature is given by

\begin{equation}
T = \left( T_d^4 +  T_{effp}^4 ,
\right)^{1/4}. \label{sbc}
\end{equation}
where
$ T_{effp}^4  = 
 3\tau_L L/( 4 \pi ac r_L^2),$ 
 with $L$ denoting the total luminosity escaping from the surface.

\noindent Here we approximate the additional optical depth above the
protoplanet atmosphere, through which radiation passes, by (Papaloizou \& Terquem 1999)

\begin{equation}
\tau_L = \kappa \left( \rho_d, T_d \right) \rho_d r_L . \end{equation}

\noindent This expresses the fact that $T$ must exceed 
$T_d$ at $r= r_L$ in order that
the luminosity be radiated away from the protoplanet into the surrounding disk.
In practice it is found for the models here that $T$  always only slightly 
exceeds $T_d$ -- i.e. $T_{effp}$ is effectively small
at $r= r_L$ (see figures \ref{fig2} and \ref{fig3} below ).

\noindent For most models we adopt  disk parameters  appropriate to 5AU
from the disk  model of Papaloizou \& Terquem (1999) with Shakura \& Sunyaev (1973) $\alpha = 0.001$
and steady state accretion rate of $10^{-7} M_{\oplus}/yr.$ Accordingly
$T_d = 140.047K$ and  $P_d = 0.131$ dyn cm$^{-2}$.
\subsubsection{Models with Free Boundary Accreting from the Protostellar Disk}
In contrast to embedded models of type A, we can consider models that have 
boundaries detatched from and interior to the Roche lobe which still accrete material
from the external protoplanetary disk that orbits the central star.
This is expected as numerical simulations of disk planet interactions have shown
that once it becomes massive enough a protoplanet forms a gap in the disk
but is still able to accrete from it through a circumplanetary disk
(see eg. Kley 1999; Nelson et al. 2000; Lubow, Siebert \& Artymowicz 1999). 
We thus consider models with free boundaries which are able to increase
their mass and liberate gravitational energy through its settling.
We subsequently refer to such models as type B.

\begin{figure*}
\epsfig{file = 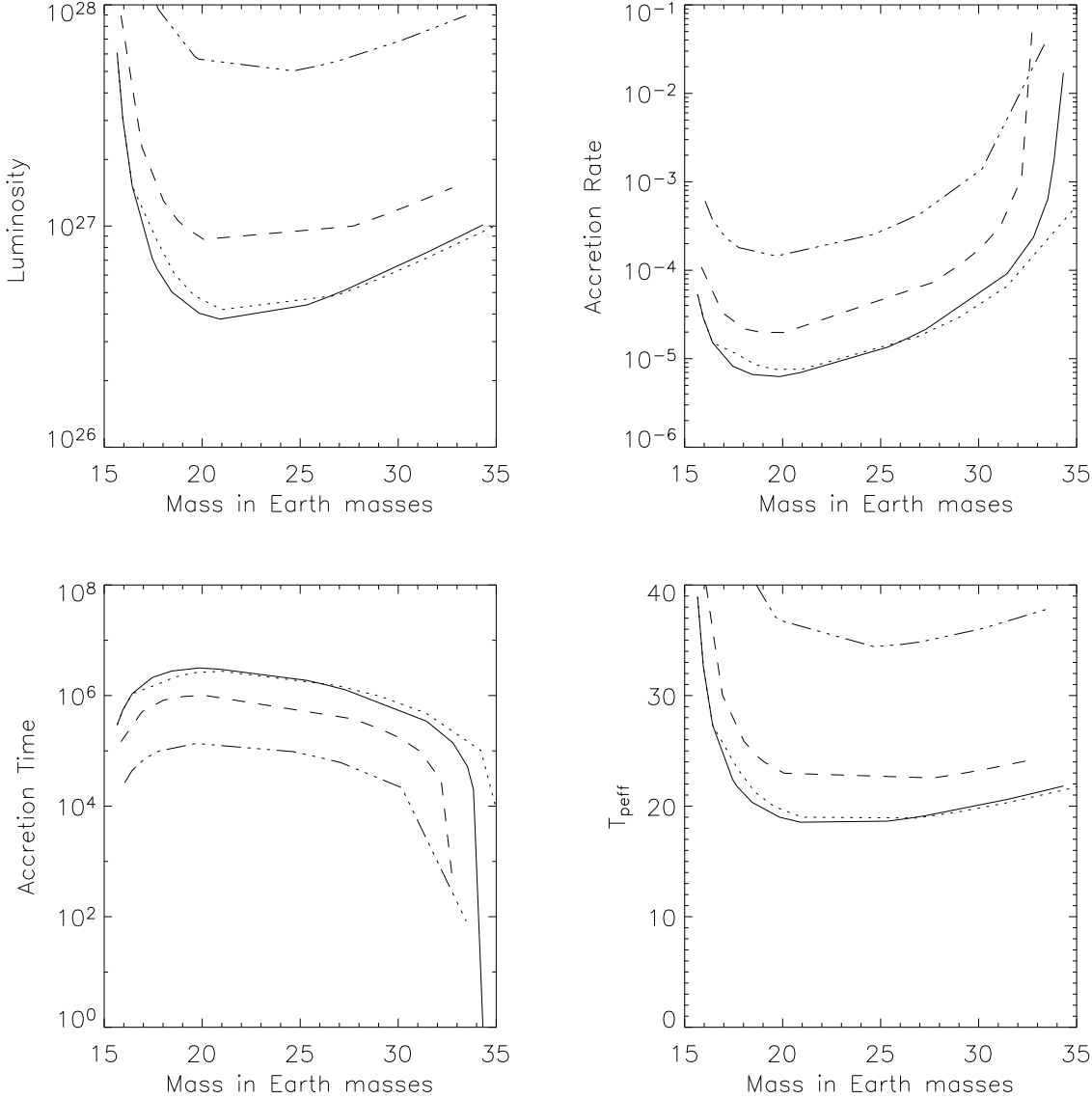, height=16cm,  width=16cm, angle=360}
\caption[]{As in figure \ref{fig2} but for two models with standard
opacity and solid cores of  $15M_{\oplus}.$
The model illustrated with the full curve is embedded
in a standard disk while the model illustrated with a dotted curve
is embedded in a disk with the same temperature
but with a density ten times larger.
These two protoplanet models show very similar behaviour
indicating lack of sensitivity to the detailed boundary
conditions.  In addition we illustrate two models with this core
mass embedded in a standard disk  but with opacities which  have a reduction factors of ten and one hundred
 (dashed curves and triple dot dashed curves respectively) that is  constant for $T <1600K$
and which then decreases linearly to unity at $T = 1700K.$}
\label{fig3}
\end{figure*}

\begin{figure*}
\epsfig{file=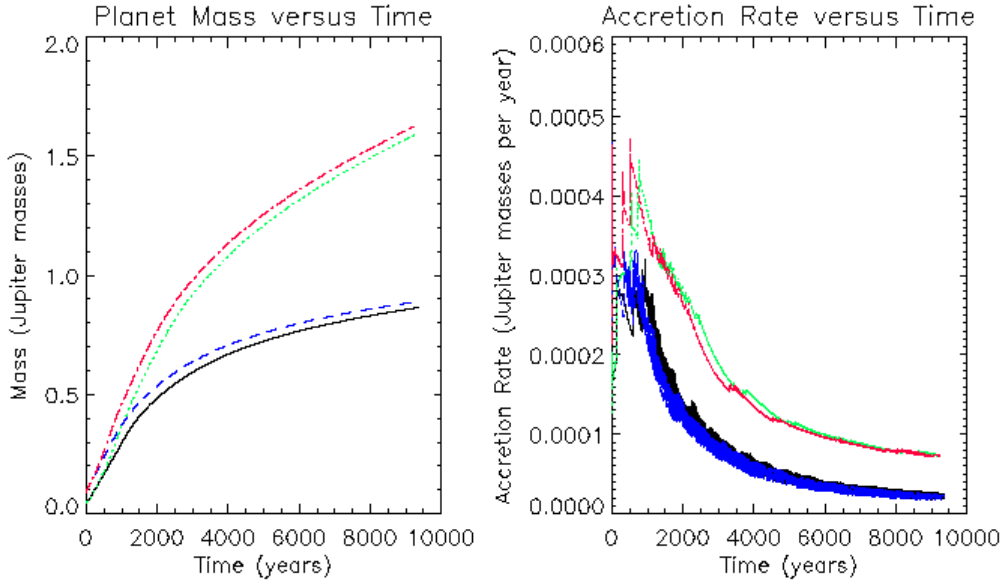,width=14cm}
\caption{This figure shows the mass accretion onto planetary cores
obtained from the hydrodynamic simulations described in the text.
The left panel shows accumulated mass onto each planet, and the
right hand panel shows the mass accretion rate in units of
Jupiter masses per year. We note that the values obtained for this quantity
span the range of values used for the accretion rates onto the
detached planet models described in section~\ref{sec:typeB-results}.}
\label{fig3a}
\end{figure*}

\begin{figure*}
\epsfig{file = 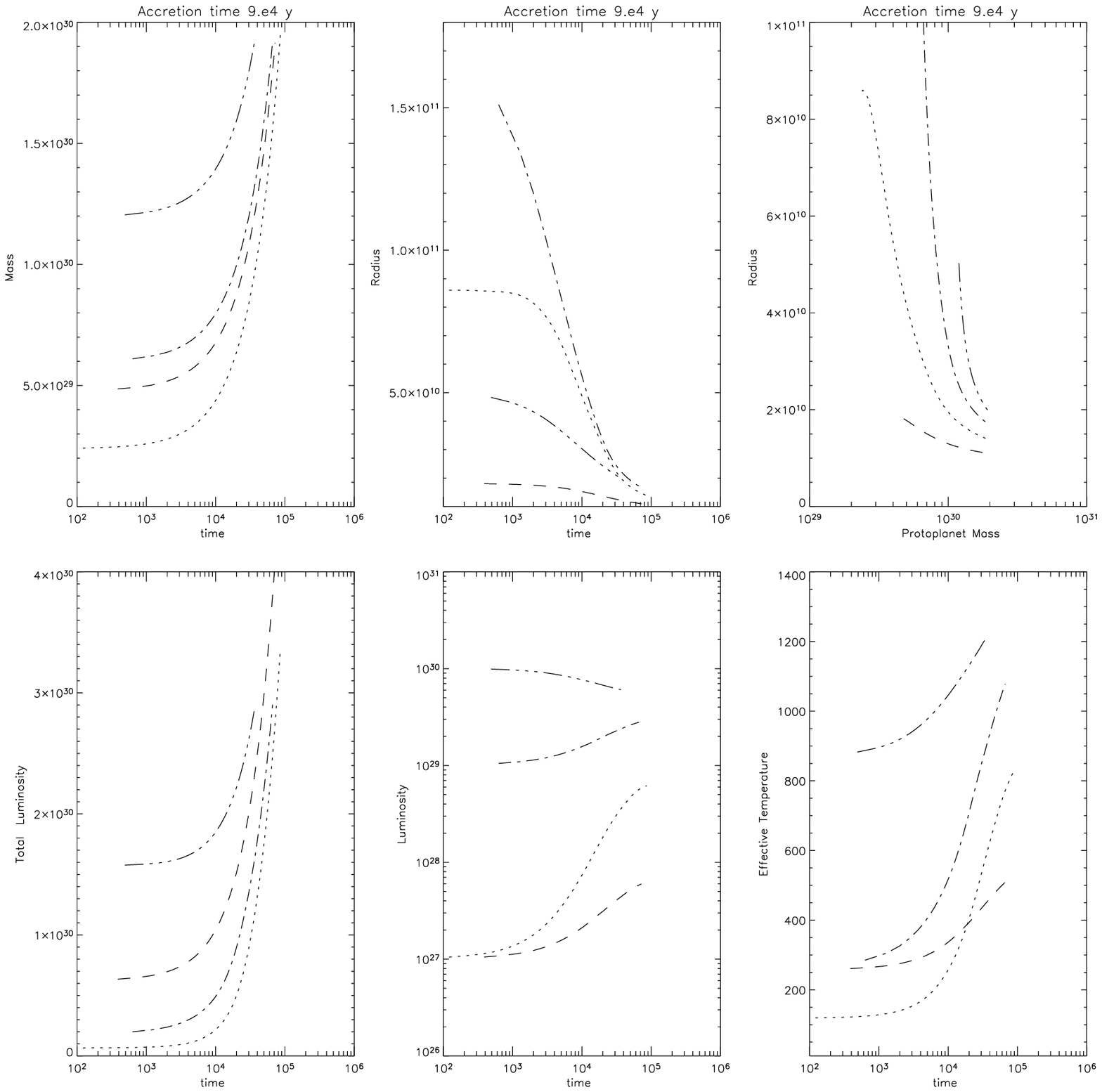, height=12cm,  width=16cm, angle=360}
\caption[]{This figure illustrates the  evolution of
protoplanet models which  accrete from the
protoplanetary disk  at an assumed
rate of one Jupiter mass in $9\times 10^4y$
but are detached from  their Roche lobes.
They have fixed
solid core masses  of $15 M_{\oplus} $ and are situated at $5AU.$
The upper left panel shows  Mass in cgs units
 as a function of time in $yr$.
 The upper central panel shows
the protoplanet radius in cgs units
as a function of time.
The upper right panel shows the  protoplanet radius 
as a function of the total current protoplanet mass.
 The lower left panel gives the total intrinsic luminosity
of the protoplanet  together with a contribution $0.5GM_{pl}/r_s(dM_{pl}/dt)$
which could be due to,  either the inner regions of the
circumplanetary disk or the disk protoplanet boundary layer
assuming small protoplanet rotation, in cgs units
as a function of time.  
The  lower  central panel gives the total intrinsic  luminosity
of the protoplanet assuming  no  contribution from the  circumplanetary
disk or
disk-protoplanet 
boundary layer.
The lower
right panel gives the effective  temperature  as a function of time.
The four models shown  correspond to differing initial conditions
corresponding to different  starting masses and luminosities.
 The same line type in different panels corresponds to the same model.
The resulting evolutionary tracks tend to show convergence
as time progresses.}
\label{fig4}
\end{figure*}

\noindent For these models, in contrast to those of type A,
the effect of the  exterior disk material  on the 
surface boundaries is small. Thus for the boundary condition on $T$
we again adopt
equation (\ref{sbc})  but with $\tau_L = 0.5$.  We note for these models $T_{effp}$ is
in general significantly larger than $T_d.$ 
For the boundary condition on $P$ we adopt
\begin{equation}
 P = P_d +{g \over \kappa}, \label{fpbc}
\end{equation}
which is the conventional stellar structure boundary condition (eg. Schwarzschild 1958)
but with the addition of the  background pressure $P_d,$
 which in fact for these models makes only a small contribution.

\noindent In order to have a complete system for which the
evolution can be calculated equations (\ref{dpdvarpi})- (\ref{mixl})
 need to be supplemented by
an equation governing  internal energy production and the internal
luminosity, normally the  first law of thermodynamics.
Here we simplify matters by using the fact that most of the internal energy
of the models is contained within a deep convection zone.
The thermal time scale associated with relaxation of the  exterior
layers is expected to be much shorter than the thermal relaxation  time scale
associated with the model as a whole. Under these conditions,
if the model evolution time scale is on the global Kelvin-Helmholtz time scale
or longer, it is a reasonable assumption that $L_r$ is constant in the outer
layers. Because of efficient convection the inner convection zone
is  unaffected by the distribution of $L_r.$ Accordingly we make the assumption
that $L_r = L $ is constant in the outer layers. This is expected to hold
during the longest lasting evolutionary phases
for all masses  and and at all times
for the
larger masses which tend to have only a very thin surface radiative shell,
but we bear in mind that it may fail when the evolution time becomes very short.
With the above assumption we  obtain a complete system for
which the evolution may be calculated.

\section{Accretion,  Settling and Evolutionary Sequences}
\label{sec:Accretion}
The models we consider here  provide an evolutionary sequence
with their mass $M_{pl}$ increasing through accretion from the protoplanetary disk.
This material also liberates gravitational energy as it settles.
To describe their evolution we consider the total energy of the protoplanet
within the Roche lobe
\begin{equation}
 E = \int_0^{M_{pl}} \left(U - {GM\over r}\right)dM.\label{toten}\end{equation}
Here $U$ is the internal energy per unit mass
and  we neglect the energy involved in bringing material from $\infty$
to the Roche lobe. This is justified because most of the mass is concentrated 
well inside it where the specific energies are much higher.

\subsection{Models of Type A}
\label{sec:typeA-evol}
\noindent We now consider models of type A. From the theory of stellar
structure,
if the source of energy
was specified,  a model would be uniquely determined once $M_{pl}$
is specified. Not specifying the source
of energy leaves one free parameter. However,
because
the radius is specified as a function of $M_{pl}$
through the Roche lobe condition, this freedom is lost so that the
models do 
form a one parameter family specified by $M_{pl}.$
Accordingly we may write $E = E (M_{pl}).$
 
\noindent When such a model increases  its  mass slightly so that 
$M_{pl} \rightarrow M_{pl} + dM_{pl},$ the change of energy content is
$dE = (dE/dM_{pl})dM_{pl}.$ 
If the energy change balances losses by radiation in time dt, $Ldt,$
then conservation of energy requires that

\begin{equation}
{dE\over d M_{pl}}{dM_{pl}\over dt} = -L,\label{eveA}
\end{equation}
with $L$ being the luminosity at the surface. This determines the evolution 
of models of type A.

\subsection{Models of Type B}
\label{sec:typeB-evol}
\noindent In contrast to models of type A, for
an assumed externally supplied accretion rate, models of type B
form a two parameter family
in that, without specification of the energy source, and given their
freedom to determine their own radius,
they require specification of both  $M_{pl}$ and
$L$ in order for a model to be constructed. Thus $E = E(M_{pl}, L).$
Accordingly for small changes in mass, and luminosity
the change in $E$ is
\be dE = \left( {\partial E\over \partial M_{pl}}\right)dM_{pl} +
 \left( {\partial E\over \partial L}\right )dL.\ee

\noindent  Now  for these  models, matter is presumed to join the protoplanet
on its equator after having accreted through a circumplanetary disk.
In this case we assume the accretion rate to be prescribed
by the dynamics of the disk--planet interaction  while 
gap formation is taking place. This is 
found to be the case  from simulations of disk--planet
interactions where it is found that an amount of material comparable
to that flowing through the disk may be supplied to the protoplanet
(Kley 1999; Nelson et al. 2000; Lubow Siebert \& Artymowicz 1999; and
simulations  presented in section~\ref{sec:hydro-sim}).

\noindent In arriving there all available gravitational binding  energy of
$-GM_{pl}/r_s$ per unit mass, $r_s$ being the surface radius,
has been liberated and so  
an amount of energy  $-GM_{pl} dM_{pl}/r_s$
must be subtracted  
from $dE$ in order to obtain the energy available to replace radiation losses.

\noindent Therefore if the changes occur over an interval $dt,$ we must have
$dE +GM_{pl} dM_{pl}/r_s=
(\partial E/\partial M_{pl})dM_{pl} + (\partial E/\partial L)dL +GM_{pl} dM_{pl}/r_s = -Ldt.$

\noindent   Thus total energy conservation
for models of type B  enables the calculation of evolutionary tracks
through.
\begin{equation}
\left[{\partial E\over \partial M_{pl}} +  {GM_{pl}\over r_s} \right ]
{dM_{pl}\over dt} +  {\partial E\over \partial L } {dL\over dt} = -L.
\label{eveB}
\end{equation}

\noindent Note that as we regard the accretion rate  
$dM_{pl}/ dt$ as specified for these models,
equation (\ref{eveB}) enables the evolution of $L$ to be calculated.

\noindent  Thus equations (\ref{eveA}) and (\ref{eveB}) constitute the basic equations
governing the evolution of models of type A and type B respectively.

\noindent 
 Note that we neglect any input from planetesimal accretion during and
after the phase when the core becomes critical. The primary reason for doing
this is that we are interested in examining the fastest time scales possible
for giant planet formation {\em via} the core instability scenario.
The inclusion of planetesimal accretion and the associated accretion
luminosity will have the effect of lengthening the time scale of formation,
provided that the core mass itself does not increase significantly.
However, there are also uncertainties about how large the planetesimal
accretion rate ought to be.

Previous work on the formation of gas giant planets {\em via} the core
instability model assumed that core formation can proceed through runaway growth
in which a protoplanetary core can grow by accreting essentially all
planetesimals in its feeding zone (e.g. Pollack et al. 1996). This resulted
in a core formation time of $\sim$ few $\times 10^5$ years. Simulations by
Ida \& Makino (1993) indicate, however, that runaway growth slows down prior
to the completion of core formation, and proceeds through a more
orderly mode of planetesimal accretion known as oligarchic growth.
This arises because neighbouring planetary embryos stir up the random motions
of the planetesimal swarm, reducing the effectiveness of gravitational focusing.
N-body simulations of protoplanetary core formation indicate that obtaining
cores of the necessary mass is not an easy task to achieve during the
oligarchic growth phase, in part due to planetary cores of a few
M$_{\oplus}$ repelling the surrounding planetesimals and opening gaps in the
planetesimal disk, and in part due to the excitation of planetesimal
eccentricities and inclinations by the `oligarchs' (e.g. Thommes, Duncan \&
Levison [2003]).

After core formation, and during the longest phase of evolution involving gas
settling onto the core,
the calculations of Pollack et al. (1996) result in planetesimal accretion
rates that are a factor of $\sim 3$ times smaller than
those of the gas accretion
rate, and this planetesimal accretion results in significant accretion
luminosity.
This planetesimal accretion arises because the feeding zone expands as
the planet mass increases due to gas accretion, and depends on the strict
assumption
that planetesimals are not allowed to enter or leave the feeding zone.
Thus the possibility of gap formation in the planetesimal disk, as found
by Thommes et al. (2003), is not accounted for in these models, although
one may reasonably expect its effect to be increasingly important
as the planet mass increases.

The generation of significant luminosity from planetesimal accretion
depends on where it is assumed the energy is deposited within the
protoplanet. Large (100 km) planetesimals are able to penetrate deep
into the planetary interior, and so provide a significant source of energy by
virtue of descending deep into the gravitational potential well. Smaller
planetesimals or fragments are more likely to dissolve higher up in
the planet atmosphere, and so will contribute less accretion luminosity.
A possible resolution of the long time scales of formation for
planetary cores reported by Thommes et al. (2003) is that collisions
between planetesimals result in fragmentation when their random motions are
excited by the forming planetary embryos (e.g. Rafikov 2004). This
possible generation of smaller planetesimals results in increased
efficiency of gas drag by the nebula in damping random motions, thus speeding up
planetesimal accretion by planetary embryos. This potential modification of
the size distribution will also have an impact on the accretion
luminosity generated by accreted planetesimals.

In the light of these uncertainties in the radial distribution and size
distribution of planetesimals, and its effect on planetesimal accretion rates
during the gas settling, and rapid gas accretion phase of giant
planet formation, we believe it is justified to treat the planetesimal accretion
rate and its associated luminosity generation as a free parameter of
the problem. A similar approach has been argued for by Ikoma, Nakazawa \&
Emori (2000). As we are interested in examining the shortest possible
time scales for giant planet formation, we neglect the effects of planetesimal
accretion in this study.

\begin{figure*}
\epsfig{ file = 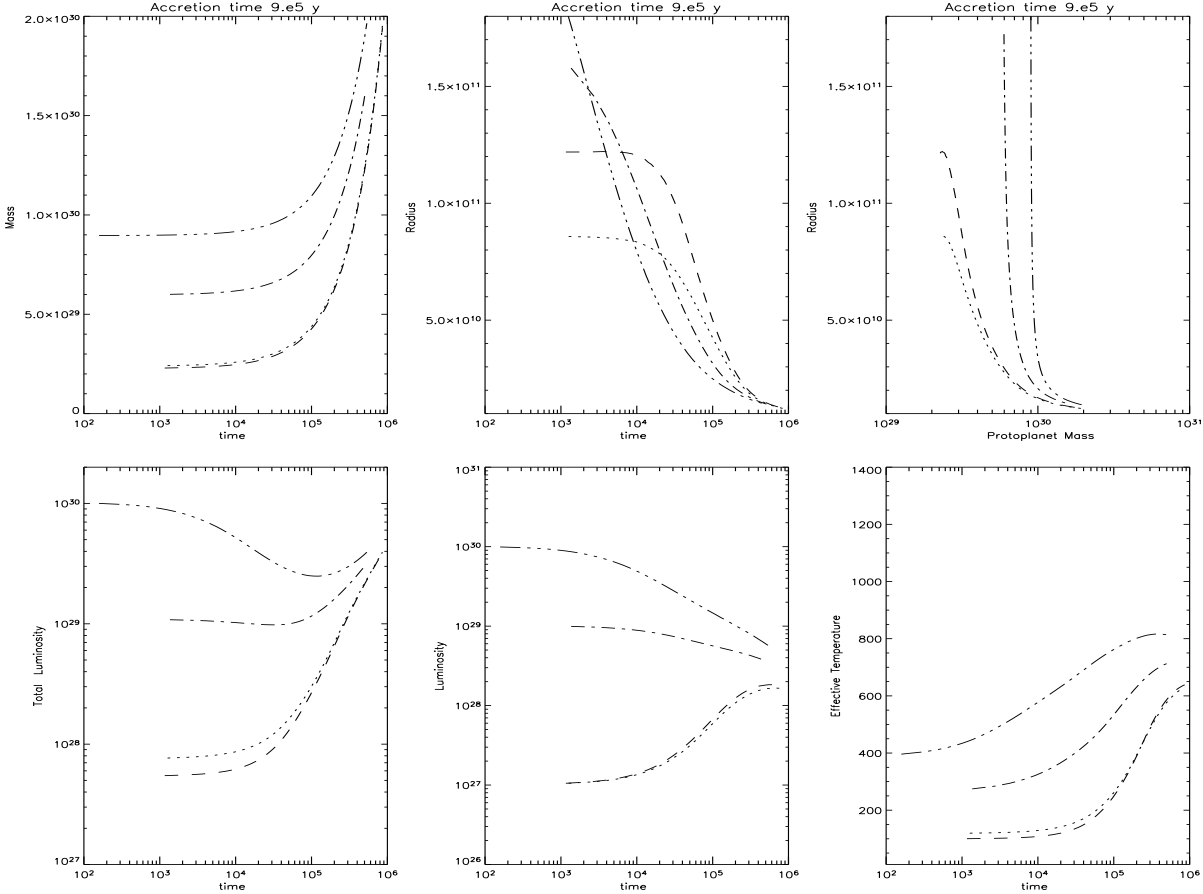, height=12cm,  width=16cm, angle=360}
\caption[]{As in figure \ref{fig4} but  for an assumed accretion rate from
the disk that is ten times slower.
As a consequence of that it takes about ten times
longer to attain one Jupiter mass in these cases. }
\label{fig5}

\end{figure*}
\begin{figure*}
\epsfig{file = 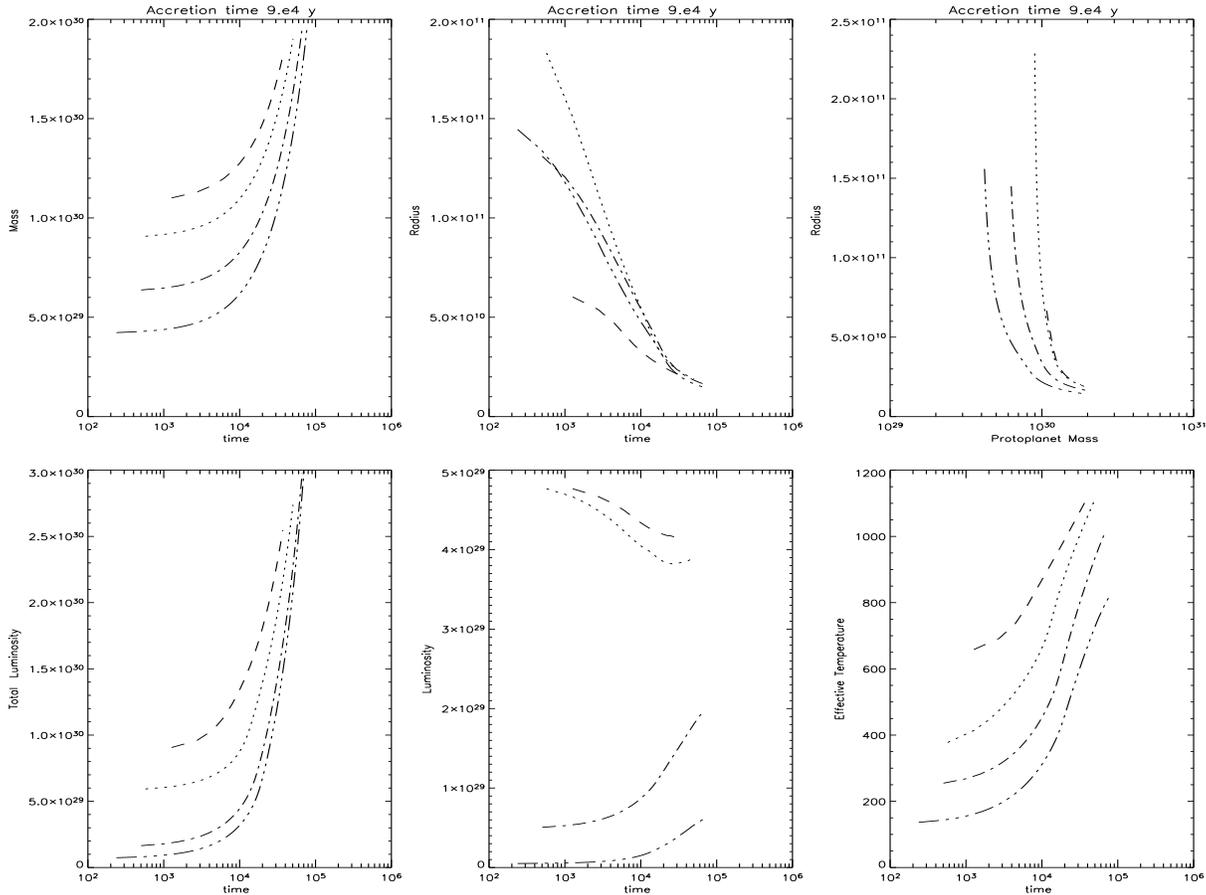, height=12cm,  width=16cm, angle=360}
\caption[]{As in figure \ref{fig4}  but for models with a $5M_{\oplus}$
solid core.}
\label{fig6}
\end{figure*}

\begin{figure*}
\epsfig{ file = 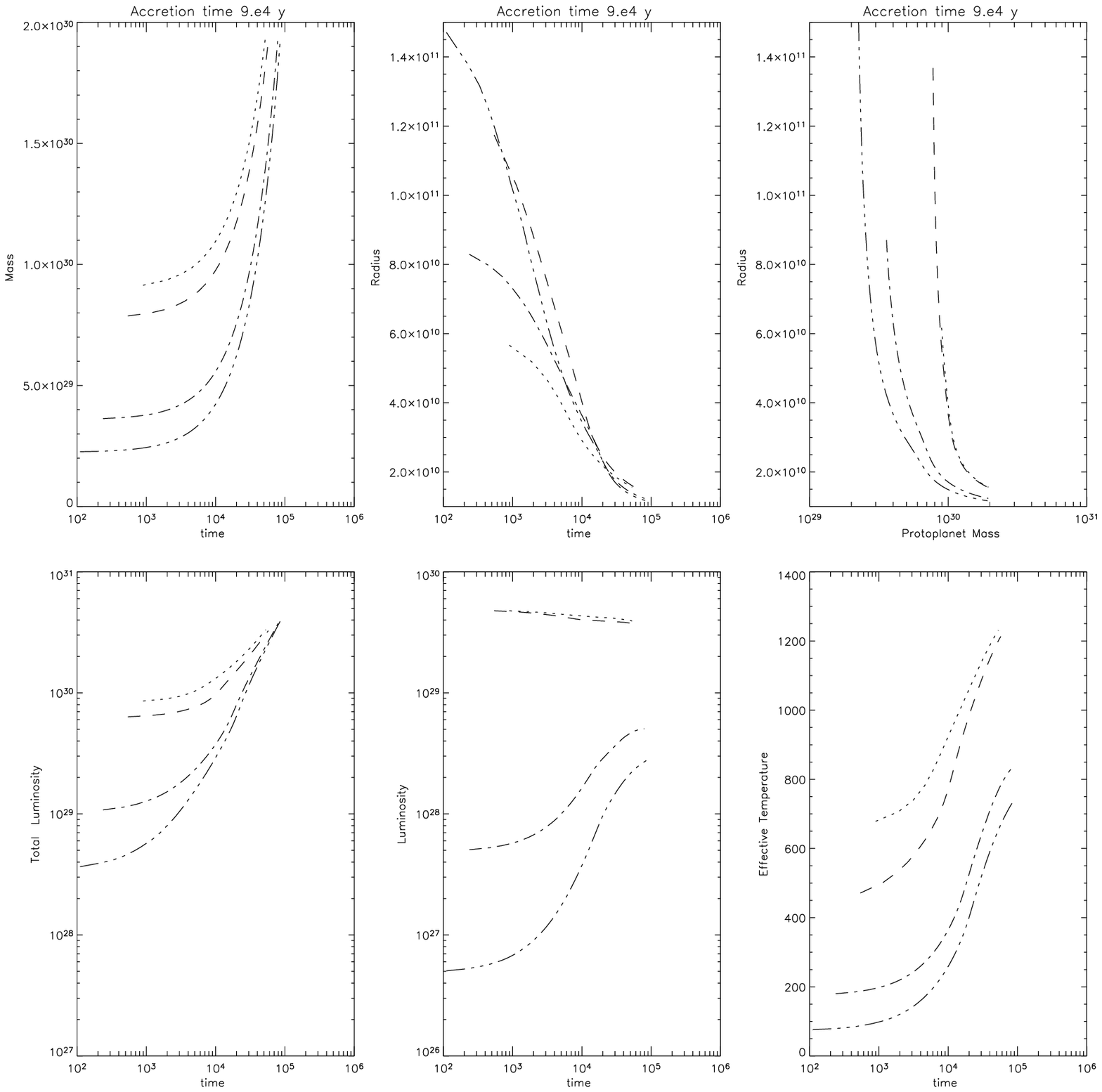 , height=12cm,  width=16cm, angle=360}
\caption[]{ As in figure \ref{fig4}  but for models with a $5M_{\oplus}$
solid core which have  opacities reduced by a factor of three.}
\label{fig7}
\end{figure*}

\begin{figure*}
\epsfig{file =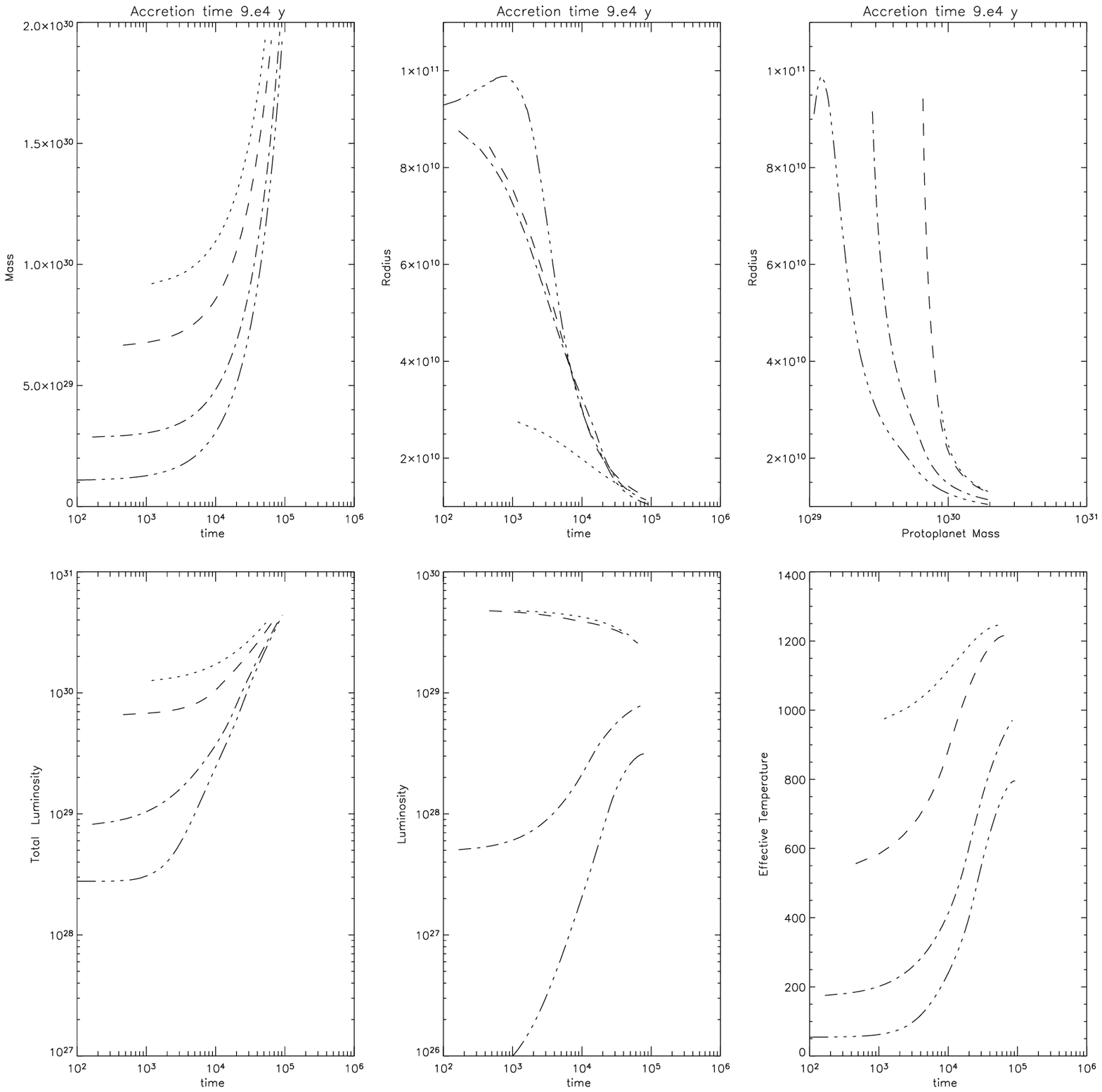, height=12cm,  width=16cm, angle=360}
\caption[]{ As in figure \ref{fig4}  but for models with a $5M_{\oplus}$
solid core which have  opacities reduced by a factor of ten. }
\label{fig8}
\end{figure*}

\noindent With the above assumption we have a complete system of equations
(\ref{dpdvarpi}) -- (\ref{mixl}), (\ref{eveA}) and (\ref{eveB}) for
which the evolution may be calculated.

\section{Disk--Planet Simulations}
\label{sec:disk-planet}
In order to estimate the rate at which an accreting protoplanet
can accrete gas from a protoplanetary disk, we performed hydrodynamic
simulations of low mass protoplanetary cores embedded in viscous
disk models. These simulations were performed with a modified version
of the grid based hydrodynamics code NIRVANA (Ziegler \& Yorke 1996).

\subsection{Initial Setup and Boundary Conditions}
The disk models are simple 2-D models with the initial
surface density given by a power law $\Sigma(R)=\Sigma_0 R^{-1}$.
We set $\Sigma_0$ such that there are 0.02 $M_{\odot}$
interior to 40 AU, similar to the minimum mass solar nebula model.
We assume a locally isothermal equation of state, and specify that
the disk vertical thickness to radius ratio have a constant value 
$H/R=0.05$. We model the angular momentum transport processes
in the disk using a simple `alpha' prescription for the disk viscosity 
in the Navier--Stokes equation -- i.e. the kinematic viscosity
is given by $\nu = \alpha c_s H$ where $\alpha$ is a parameter,
$c_s$ is the sound speed, and $H$ is the local disk thickness.
We consider values of $\alpha=5 \times 10^{-3}$ and $10^{-3}$.

The number of grid cells used was $(N_R,N_{\phi})=(260,630)$.
The inner boundary of the computational domain was placed at
$R=0.4$ and the outer boundary at $R=3$. Reflecting boundary conditions
were used at both radial boundaries, and linear viscosity was used between
$0.4 \le R \le 0.6$ and $2.5 \le R \le 3$ to reduce reflection of waves excited by
the planet. The gravitational potential of the planet was softened using a
softening parameter $b=0.5 H(R_p)$ -- i.e. half of the local disk 
semi--thickness.

Simulations were initiated by placing a low mass planet (either 15 or 30 Earth
masses) at a radius $R_p=1$ in the disk. The planet was assumed  to accrete gas
that entered its Hill sphere. This was achieved by removing gas
from any cells that lay within half of the planet Hill sphere. The e-folding
time of this gas removal was $\tau_{acc}=\Omega^{-1}$, where
$\Omega=\sqrt{(G M_*)/R_p^3}$. Thus this corresponds
to the extreme case when the planet accretes material within
the Hill sphere on the dynamical time scale. The gas that was removed from the Hill sphere
was added to the planet at each time step, such that the planet mass
is a function of time. Similar models are described in Nelson et al. (2000).

\section{Protoplanet Model Calculations}
\label{sec:envelope_calc}
We solve  equations~(\ref{dpdvarpi}-\ref{mixl})
with the boundary conditions described above to
get the structure of the  protoplanet models.  

\noindent For a fixed accretion rate onto a core
$\dot{M}_{core}$ at a given radius, there is a critical
core mass $M_{crit}$ above which no solution can be found
in hydrostatic and thermal equilibrium that joins on to
the protoplanetary disk model assumed at the Roche lobe.

\noindent In this paper we consider cores with 
$ M_{core} = 15M_{\oplus}$ and $ M_{core} = 5M_{\oplus}.$
Our evolutionary calculations commence close to the state when the cores
are critical, that is no further gas 
can be added in strict hydrostatic equilibrium. At this stage the evolution
is slowest and we compute a type A model sequence by use of equation
(\ref{eveA}). These models are in contact with the Roche lobe.

\noindent We also construct type B model sequences.
These satisfy the free surface boundary conditions given by
equations (\ref{sbc}) and (\ref{fpbc}). Because the surface is free,
these form a two parameter sequence in that evolutionary tracks
for a given mass can be started for a range of radii
(or equivalently the luminosity may be used as a parameter).
This the same situation as in standard pre--main sequence contraction
where a stellar model of a given mass can be started at different points
on an evolutionary track corresponding to different radii.

\noindent  We have considered models using the Bell \& Lin (1994) opacities
hereafter referred to as standard.
These have a very large contribution from dust grains for $T < 1600K$
and because there is clearly some uncertainty about the disposition of the dust
particularly under circumstances where the protoplanet is assumed isolated
from further planetesimal accretion, we have explored the effect of 
reducing this contribution to the opacity by factors of up to $100$
for both models of type A and B. We have done this, by making the reduction for
the opacity as a whole, for
$T <1600K$ only, and  with a reduction factor that is  constant for $T <1600K$
and which then decreases linearly to unity at $T = 1700K.$
In practice we find that the results are essentially  independent of whether
such a linear join is made or not. The  
uncertainty in the magnitude
of the surface opacity  as well as its important role in 
controlling the evolutionary time scale
of an embedded protoplanet has been pointed out by
Ikoma Nakazawa \& Emori (2000).

\subsection{Models of Type A}
\label{sec:typeA-results}

\noindent We begin by describing some typical models of type A.
In figure \ref{fig0}, state variables are plotted for a protoplanet model
with $ M_{core} = 15M_{\oplus}$ which has 
a total mass $ 25.3 M_{\oplus}.$ As expected the deep interior
of this model is convective with little energy transported by radiation.
However, there are two convective regions which occur for
$679K > T > 263K$ and $T > 2100K.$
Approximately ninety  eight percent of the mass
is in the inner  convective zone. This means that most of the thermal inertia
is contained within the deep convection zone rendering the assumption
of little spatial variation of the luminosity in the upper layers a reasonable
approximation. The existence of two separate convective regions
is in contrast to what we find for models of type B that approach $1M_J.$
In those cases we find a single interior convection zone for $T > 1000-2000K,$
with negligible mass in the outer radiative region.

\begin{figure*}
\epsfig{file = 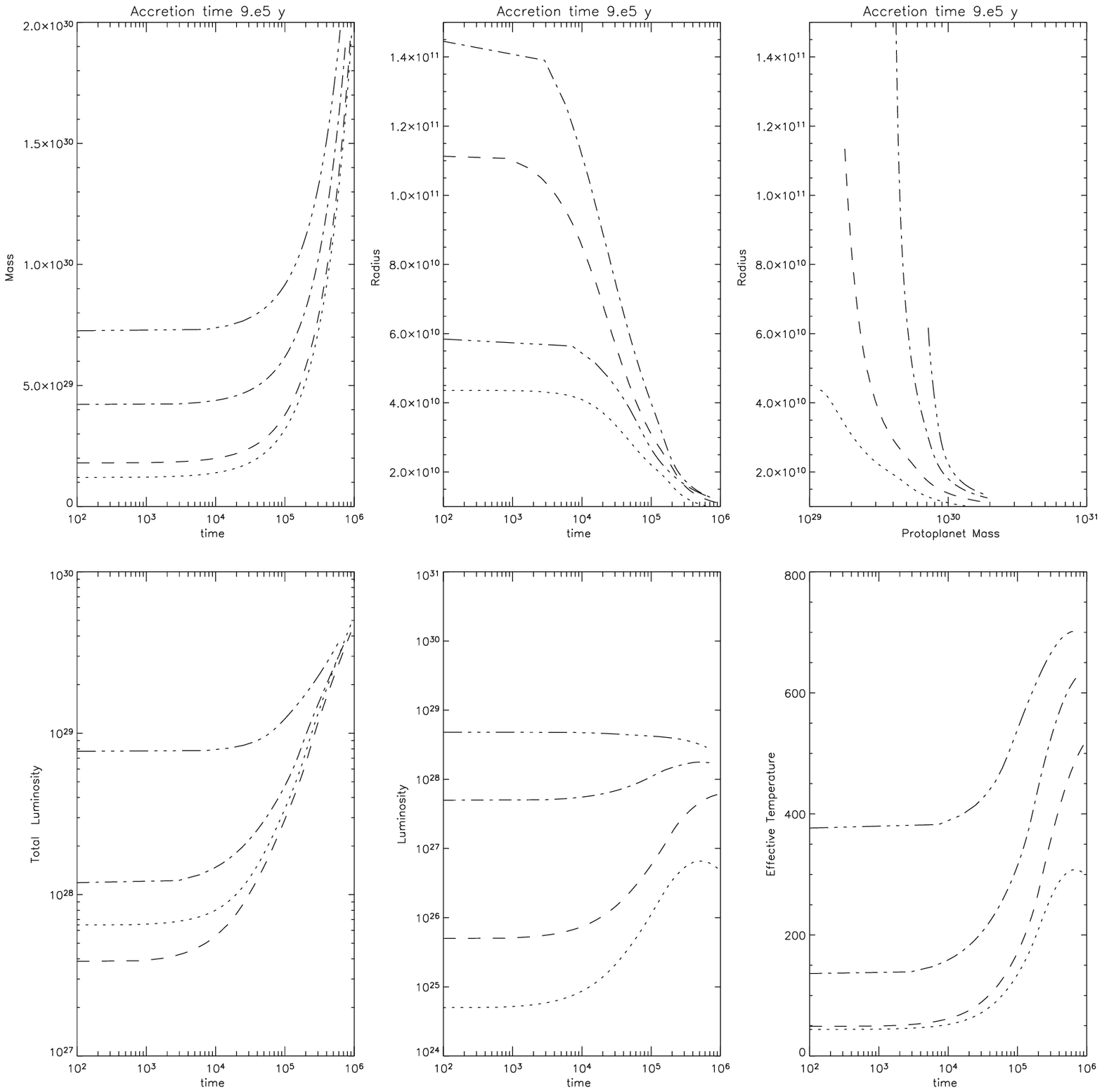, height=12cm,  width=16cm, angle=360}
\caption[]{  As in figure \ref{fig4}  but for models with a $5M_{\oplus}$
solid core accreting from the disk at a rate that is ten times slower.
As a consequence of that it takes about ten times
longer to attain one Jupiter mass.}
\label{fig9}
\end{figure*}

\begin{figure*}
\epsfig{ file =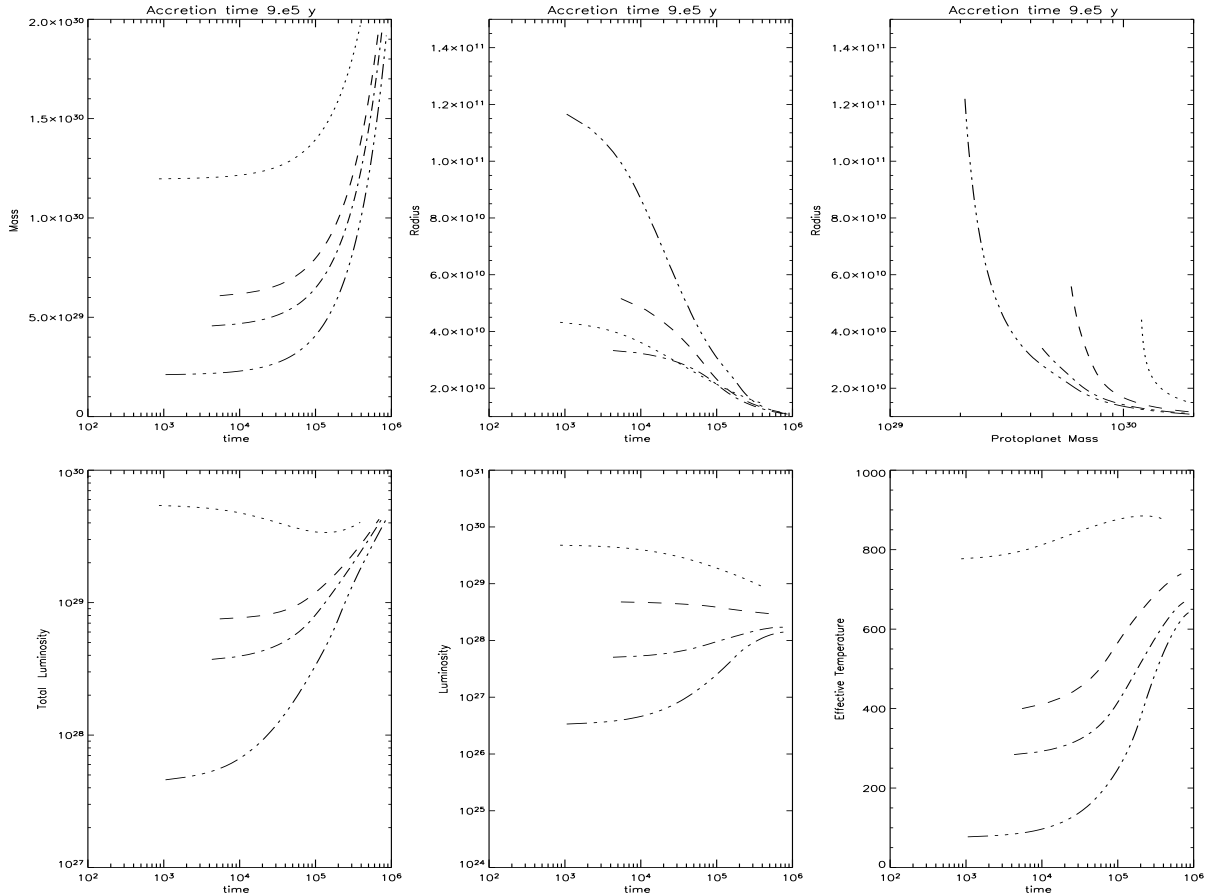, height=12cm,  width=16cm, angle=360}
\caption[]{ As in figure \ref{fig4}  but for models with a $5M_{\oplus}$
solid core accreting from the disk at a rate that is ten times slower
and which have opacities a factor of three smaller.}
\label{fig10}
\end{figure*}

\begin{figure*}
\epsfig{ file =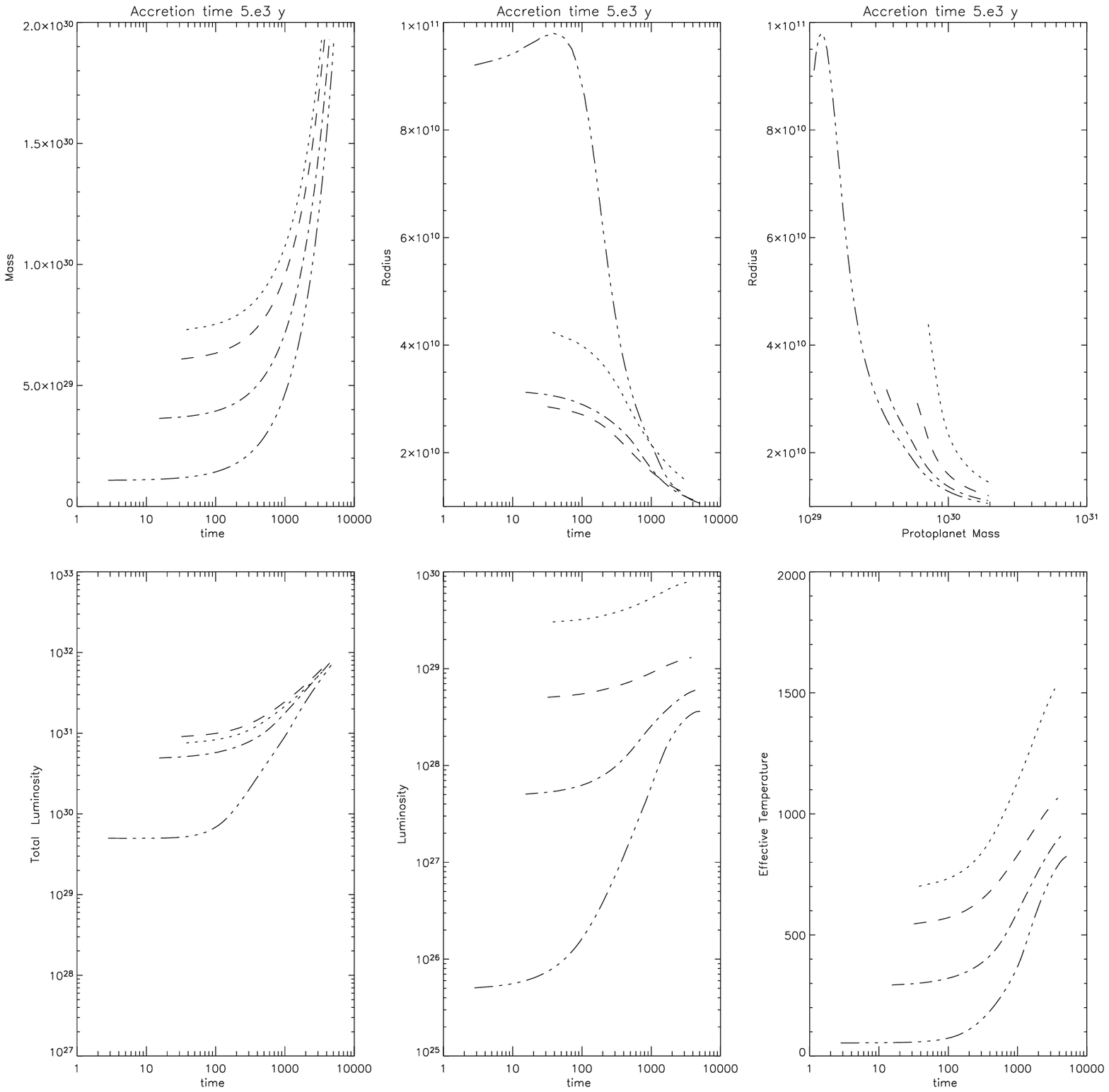, height=12cm,  width=16cm, angle=360}
\caption[]{ As in figure \ref{fig4}  but for models with a $5M_{\oplus}$
solid core accreting from the disc at a rate that is  $18$ times  faster
and which have opacities a factor of  ten smaller.}
\label{fig11}
\end{figure*}

\noindent In figure \ref{fig1},
we illustrate the behaviour of the state variables
for a protoplanet model with the smaller $5M_{\oplus}$ core mass.
The total mass is $17.6 M_{\oplus}.$  This has similar
properties to the previous case regardless of the 
fact that the core mass is three times smaller.
Convective  heat
transport occurs when
$720K > T  > 264K $ and when $T > 2100K. $
The inner eighty percent of the mass
is convective.

\noindent In figure \ref{fig2} we illustrate the evolution of
models of type A for 
$ M_{core} = 5 M_{\oplus}.$
Cases with standard opacities and with  opacity reductions of
three, ten and one hundred made globally and for $T < 1600K$
are shown.
In all cases as the models gain in mass from the protoplanetary disk
their luminosity  increases and their evolutionary time
measured through their accretion time $M_{pl}/(d M_{pl}/ dt)$
decreases. In the standard opacity case the accretion 
time is very long, exceeding $10^8y.$ However, this time
reduces by the opacity reduction factor independently 
of where this is applied even though the situation
might have appeared to have been complicated by the existence of two
convective regions.  
Thus times $\sim 3\times 10^6y$ are attained 
for reduction factors of one hundred.

\noindent The evolutionary time scale of the models begins
to decrease rapidly once $M_{pl} \sim 20M_{\oplus}$,
becoming less that $10^5y$ even for the standard opacity case.
This phenomenon, which is sensitive to relatively minor  model
details, can be traced to the fact that $d E/dM$ becomes small 
or that less and less binding energy is liberated
as the mass increases. This is likely to indicate the 
onset of a rapid collapse and possible detachment from the Roche lobe.
As the  effect of disk planet interactions and local gas depletion
are likely to become important, we have not tried to follow
such rapid evolution with the simplistic models adopted here.
Rather we have considered evolutionary sequences
of type B which are likely to be the outcome. 
Although the position where such a sequence is joined
cannot of course 
be determined without considering the above rapid phase of evolution.

\noindent Figure \ref{fig2} also shows that 
$T_{effp},$ the effective temperature needed to
supply the luminosity of the model, is always small when compared to
the surrounding protostellar disk temperature.
This indicates negligible thermal perturbation of the protostellar disk.

\noindent In figure \ref{fig3} 
we illustrate the evolution of
models of type A for
but for two models with standard
opacity $M_{core} = 15M_{\oplus}.$
For these models the longest evolutionary times
are in the $3 \times 10^6 y$ range.
The attainment of short evolutionary times likely
leading to detachment from the Roche lobe occurs
for $M_{pl} \sim 35 M_{\oplus}$ in this case.
The two models illustrated differ in surface boundary conditions.
The model illustrated with dotted curves
is embedded
in a disk 
with the same temperature
but with a density ten times larger than usual. Except
during the beginning of the rapid evolution phase
the models show very similar behaviour.

\noindent In figure \ref{fig3} we also plot
evolutionary tracks for which the opacity
was reduced by factors of ten and one hundred
in the surface layers for which $T > 1600K$
with a linear transition to standard opacities occurring
for $1700K >  T > 1600K.$
For these sequences the transition mass is unaffected but
the evolutionary time scales are  factors of three and thirty 
faster respectively. This means that the formation time scale
is reduced to $\sim 10^5 y$ in the latter case.

\subsection{Planet Accretion Rates}
\label{sec:hydro-sim}
As described in section~\ref{sec:typeB-evol}, the planetary models that 
we construct with
a free surface require a gas accretion rate to be specified.
We have performed simulations of low mass protoplanets embedded in 
protoplanetary disks to estimate the accretion rate onto a protoplanet
that may be supplied by a protoplanetary disk, using different assumptions
about the initial planet mass and disk viscosity. The results of these
simulations are presented in figure~\ref{fig3a}. The left hand
panel shows the evolution of the planet mass,
and the right hand panel shows the accretion rate as a function
of time. We considered initial
protoplanet masses of $M_{pl}=15$ and 30 Earth masses, and viscosities
with $\alpha=10^{-3}$ and $\alpha=5 \times 10^{-3}$.
The solid line in figure~\ref{fig3a} shows the model with
$M_{pl} =15$ M$_{\oplus}$ and $\alpha=10^{-3}$. The dashed line shows the model
with $M_{pl} = 30$ M$_{\oplus}$ and $\alpha=10^{-3}$.
The dotted line shows the model with $M_{pl} =15$ M$_{\oplus}$ and 
$\alpha=5 \times 10^{-3}$, and the dot-dashed line shows the model
with $M_{pl} = 30$ M$_{\oplus}$ and $\alpha=5 \times 10^{-3}$.
It is clear that the initial mass assumed for
the protoplanet is unimportant as the models quickly converge.
However, quite differing evolutionary sequences are obtained as a function
of disk viscosity. For higher viscosity, a protoplanet that is thermodynamically
permitted to 
accrete gas rapidly from a disk can grow to become a Jupiter mass in 
around 3000 years.  For lower viscosity the
growth time can be extended considerably, with an $\alpha=10^{-3}$
requiring a time $\ge 2 \times 10^4$ years for a Jupiter mass planet to form.
The simulations presented here are too low in resolution to
model the circumplanetary disk that is expected to form around the
accreting protoplanet or its interaction with it, and in principle the requirement that material
accrete through this circumplanetary disk before reaching the planet surface 
could act as a bottle neck and significantly extend these accretion time scales.
However, simple estimates of the accretion time through
such a circumplanetary disk, and the accretion rates obtained from high resolution 3-D simulations
(e.g. D'Angelo, Kley \& Henning 2003) suggest that this is not the case.
 These high resolution simulations indicate that
the circumplanetary disk that forms within the
planetary Hill sphere has a radius $\simeq 2 R_H/3$ where
$R_H=R_p(M_{pl}/3M_*)^{1/3}$ is the Hill sphere radius.
The viscous evolution time through a disk of such radius is
$$\tau_{\nu} \simeq \left( \frac{2}{3} \right)^3 \frac{R^2_H}{\nu} $$
where $\nu = \alpha c_s H$ is the kinematic viscosity, $c_s$ being the
sound speed at the outer edge of the circumplanetary disk, and $H$ being
the disk semi--thickness there. The viscous time scale may be expressed in
units of the planet orbital period as
$$\tau_{\nu} \simeq \frac{(3.2)^{-1}}{2 \pi \alpha}  \left(\frac{2}{3}\right)^3
\left(\frac{R_H}{H}\right)^2. $$
\noindent We note that $R_H/H \equiv {\cal M}$,
where ${\cal M}$ is the Mach number of
the flow in the outer regions of the circumplanetary disk.
Simulations by D'Angelo, Henning, \& Kley (2003), that account for heating
and cooling of the circumplanetary disks, results in Mach numbers of
${\cal M} < 2$ in their outer parts, indicating that these disks
are rather thick. If we adopt the values of $\alpha$ used in the
numerical simulations, we  estimate `that
the accretion time through the circumplanetary disk is
$\tau_{\nu} < 141$ yr for $\alpha=5 \times 10^{-3}$, and
$\tau_{\nu} < 707$ yr for $\alpha=1 \times 10^{-3}$.
These time scales should be compared with the mass accretion times
presented in figure~\ref{fig3a}. For the $\alpha=5 \times 10^{-3}$
runs the mass doubling time scale is found to be $\approx 500$ yr. For
$\alpha=1 \times 10^{-3}$  this accretion time is $\approx 1000$ yr.
This suggests that the simple prescription for modeling mass accretion
in the simulations does not significantly affect the long term accretion
times presented, as the protostellar disk supplies mass to the protoplanet
on time scales longer than reasonable
accretion times through the circumplanetary disk.
We note, however, that more detailed 3D simulations including a proper
account of the thermodynamic evolution of the gas will be required
to definitively settle this question.

The accretion times obtained in figure~\ref{fig3a} range from a 
few thousand years to a few tens of thousands of years, and show that 
the actual accretion time
obtained is sensitive to the disk viscosity assumed. In the planet models of 
type B presented below, we consider accretion times of between $5 \times 10^3$
to $9 \times 10^5$ years.
These cover the accretion times obtained for a protoplanet on short time
scales  in the extreme case when it is
immersed ab initio into an unperturbed disk as in the above simulations.
They  also allow for the situation where there is gas depletion
such that the protoplanet can only accrete for longer time scales  at an assumed  mass flow
rate through the protostellar disk of $10^{-9} M_{\odot} y^{-1}.$

\subsection {Models of Type B}
\label{sec:typeB-results}
Figure \ref{fig4} illustrates the evolution of
protoplanet models which accrete from the
protoplanetary disk at an assumed fixed
rate of $1M_J$ in $9\times 10^4y.$
This accretion rate corresponds to $1.1 \times 10^{-8} M_{\odot} y^{-1}$
which is typical of T Tauri disks, but it leads to a rapid final
accretion time for a Jupiter mass if most of this
is accreted by the protoplanet as is indicated by the
simulations of disk--planet interaction presented in section~\ref{sec:hydro-sim}
(see also Kley [1999]).
The  models have 
$ M_{core} = 15 M_{\oplus}.$
For these models the duration of the evolution is determined
by the accretion rate and is terminated when the protoplanet
reaches $1M_J.$
The four models shown correspond to 
different starting masses and luminosities.
For a given mass is it is possible to start with a range of luminosities
or for the same luminosity it is possible to start with a range of masses.
Here the dashed and dot--dashed curves correspond 
to models which start with almost the 
same mass but very different luminosities
while the dotted and dashed curves start with the same luminosity
but differ in mass by a factor of $2.5.$  
The resulting evolutionary tracks tend to show convergence
as time progresses especially in the case of the radius which ends 
up in the range $2.25 \pm 0.75 \times 10^{10} cm$ 
after $\sim 2 \times 10^4 y.$ The models attain $T_{eff} > 400K$
for most of the evolution. However
the luminosity expected from circumplanetary
disk accretion  $3.26\times 10^{30}$ erg s$^{-1}$
is only approached by the most luminous model.

\begin{figure*}
\epsfig{ file =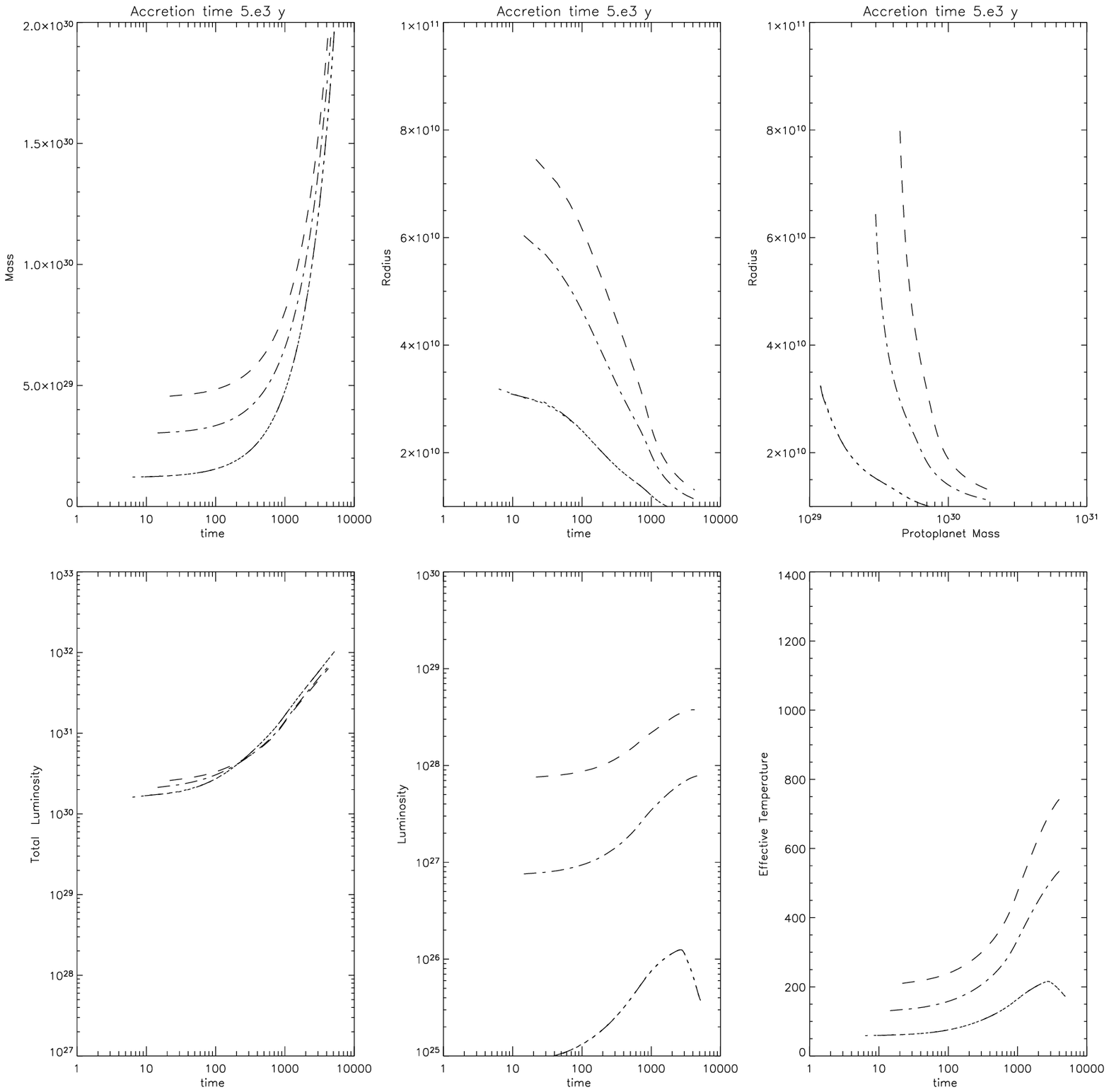, height=12cm,  width=16cm, angle=360}
\caption[]{ As in figure \ref{fig11}  but  for a sequence of models
with the opacity reduction applied only for $T<1600K$.}
\label{fig12}
\end{figure*}

\begin{figure*}
\epsfig{file=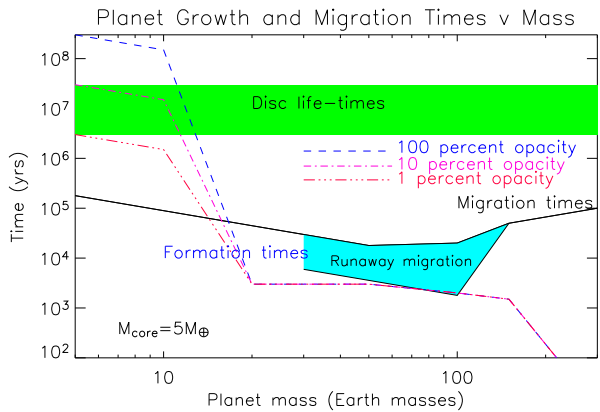, width=\columnwidth}
\caption[]{This diagram provides a schematic representation of
the formation and migration time scales of planet models as a function
of protoplanet mass for core masses of 5 M$_{\oplus}$. The range of
plausible protostellar disk life-times is indicated by the upper shaded
region spanning the times $3 \times 10^6$ -- $3 \times 10^7$ years.
The migration time as a function of planet mass is indicated by the solid
line. A shaded region indicating the `danger zone' for rapid type I
or runaway migration is also indicated. The growth time of protoplanets
as a function of planet mass is given for standard opacity (dashed line),
10 percent opacity (dashed-dotted line), and 1 percent opacity
(dashed-dot-dot-dotted line). See text for discussion of this figure.}
\label{fig13}
\end{figure*}

\begin{figure*}
\epsfig{file=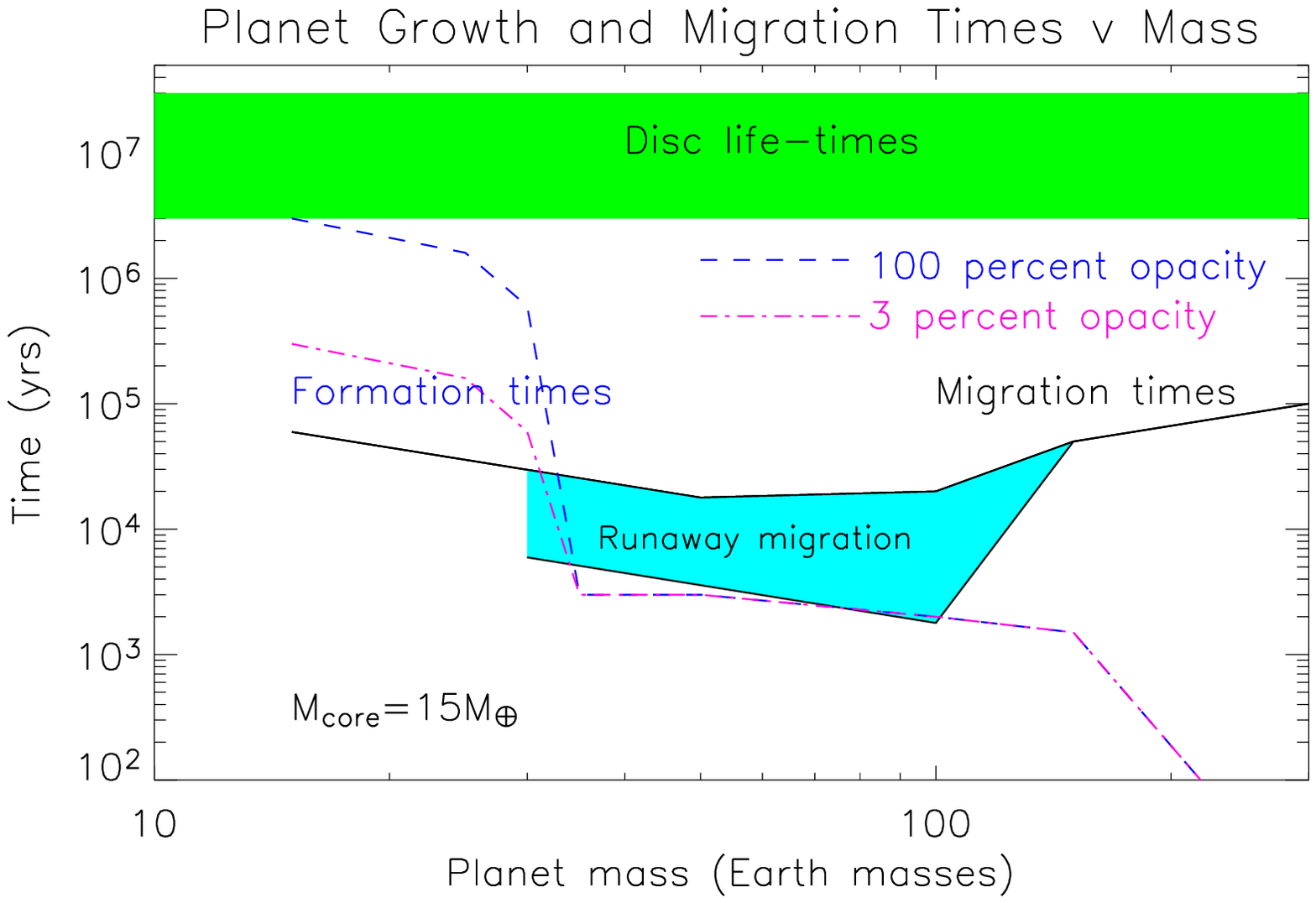, width=\columnwidth}
\caption[]{This diagram is similar to figure~\ref{fig13}, except that
it applies to planet models with 15 M$_{\oplus}$ cores. The dashed
line shows the growth time as a function of mass for models with
standard opacity. The dashed-dotted line shows the growth time
for models with 3 percent opacity. See text for discussion of this
figure.}
\label{fig14}
\end{figure*}

\noindent Figure~\ref{fig5} illustrates evolutionary tracks
for a fixed  assumed accretion rate
that is ten times slower.
As a consequence of this the evolution times are ten times
longer. The convergence of the evolutionary tracks is greater in this case
with all radii being close to $\sim 3\times 10^{10}cm$ after $\sim 10^5 y.$
The indication is that values of $T_{eff} \sim 700-800K$ 
for these models are sustained for $\sim 10^6 y.$
However, the luminosity expected from circumplanetary
disk accretion  at the later phases  $\sim 3.26\times 10^{29}$ erg s$^{-1}$
(calculated adopting a radius of $2 \times 10^{10}$ cm
for the protoplanet) is only  exceeded  at early times
by the most luminous model
which  then becomes fainter at later times. But note that for this
sequence of models and others presented later, models accreting from the 
disk can exist which have small luminosities $<<$ the circumplanetary
disk luminosity and also that due to  the protoplanet disk boundary layer,  equal to 
$0.5GM_{pl}/r_s(dM_{pl}/dt)$
(see eg. Lynden-Bell \& Pringle 1974).

\noindent In figure~\ref{fig6} we show tracks for an accretion rate
of $1M_J$ in $9\times 10^4y$ for models with  $M_{core} = 5M_{\oplus}.$
The behaviour is similar to that in the higher core mass case.
In figure \ref{fig7} we show models with $ M_{core} = 5M_{\oplus}$
with the same accretion rate
which have opacities globally reduced by a factor of three
and in figure \ref{fig8} 
the reduction is by  a factor of ten. In all of  these cases 
there is a tendency of the tracks to converge  especially
the radii of different models to a value of about $2 \times 10^{10} cm,$
 with the lower opacity models being
slightly smaller. In all cases the protoplanet luminosities
is exceeded at late stages  by the circumplanetary  disk luminosity.

\noindent In figure \ref{fig9} we illustrate models 
with $M_{core} = 5M_{\oplus}$ and standard opacities accreting 
from the disk at a rate that is ten times slower while in
figure \ref{fig10}  
the opacity is globally reduced by a factor of three at that accretion rate.
In these cases the evolution is prolonged by a factor of $10.$
These models are again similar to the previous ones and indicate that
a model starting from one Saturn mass and radius $\sim 6\times 10^{10}cm$
could sustain effective temperatures 
$\sim 700K$ for times approaching $10^6y.$

\noindent Finally in figures \ref{fig11} and \ref{fig12}  
we explore models with  $M_{core} = 5M_{\oplus}$
subjected to a very high
accretion rate from the disk at a  rate of $1M_J$
in $5 \times 10^3 y$
with an opacity reduction by a factor of ten applied globally
in the former case and applied only for $T<1600K$ in the latter.
Paradoxically (see the discussion in section~\ref{sec:discussion}) these 
models  may appear somewhat cooler
and less  intrinsically luminous than those calculated for lower accretion rates.
However, convergence of model radii towards $2\times 10^{10} cm$
again occurs.

\section{Discussion }
\label{sec:discussion}

\subsection{Planetary Growth and Migration}
In this section we discuss the evolutionary sequences associated with
planetary models of type A and B in the context of disk--planet interactions
and planetary migration. 
\subsubsection{5 M$_{\oplus}$ core models}
In figure~\ref{fig13} we have plotted a schematic
diagram showing the variation of planetary growth times as a function of
planet mass for models with cores of 5 M$_{\oplus}$.
 Here we emphasize that the growth time referred to here and below
 apply
to the gas accretion phase and not to the time  required for the solid
core to form. 
Also plotted in this diagram
is a shaded region which shows the range of T Tauri disk life--times as
inferred from infrared observations
( eg. Beckwith, Sargent, Chini \& Guesten 1990; Sicilia - Aguilar et al. 2004)
ranging between 
$3 \times 10^6$ to $3 \times 10^7$ years. We have also plotted
migration times appropriate to a standard
quiescent disk  as a function of planet mass, where the migration
rates are taken from Tanaka, Tacheuchi, \& Ward (2002). We assume that
the disk surface density scales as $\Sigma(R) \propto R^{-1}$, that
the disk surface density at 5 AU (assumed to be the planet semi major axis) is
$\Sigma(R_p)=160$ g cm$^{-2}$, and that $H/R=0.05$ (in other words the model
is similar to the minimum mass solar nebula model).  
In plotting this diagram we
also take account of the fact that more massive planets begin to open gaps,
and make a transition to type II migration (Ward 1997) , for which Jovian mass planets
migrate on the viscous time scale (here assumed to be $10^5$ years).
Also included is a shaded area for masses between 30 and 100 M$_{\oplus}$ 
that takes account the possibility of 
fast or  runaway migration for this mass range
(Masset \& Papaloizou 2003).
This   can only occur for disk masses greater than the minimum mass solar
nebula model. We make a rough estimate of  the migration
rate associated with runaway migration as being the type I rate
for a disk surface density 5 times larger than the minimum mass model.
We also assume that during runaway migration planets up to Saturn's mass
undergo migration at the appropriate type I rate, which is implied approximately
by the results of Masset \& Papaloizou (2003).

We plot the growth times for three different evolutionary models
in figure~\ref{fig13}. The dashed line represents the model with standard
opacity, the dashed-dotted line the model with one tenth the standard 
opacity, and the dashed-dot-dot-dotted line the model with opacity
reduced by a factor of 100. During the earliest phases of evolution
the growth times of these models are $\simeq 3 \times 10^8$, $3 \times
10^7$ and $3 \times 10^6$ years respectively. 
The two models with largest opacity are thus unable to form giant planets
within the disk life--time. Such systems will result in rock and ice cores
forming that are unable to accrete significant gas envelopes.

The lowest opacity model has a sufficiently
low growth time that it will be able to form a giant planet before
disk dispersal. However, figure~\ref{fig13} shows that during the early
stages of evolution, while the planet mass is below $\simeq 15$ M$_{\oplus}$,
the growth time is significantly longer than the type I migration time scale,
implying that the protoplanet will migrate into the central star before
forming a gas giant. This is a problem for all reasonable core
instability models of gas giant formation, since there exists a 
bottle neck for gas accretion while the planet mass is relatively small,
but massive enough to undergo quite rapid migration. If the core instability
model is correct, then we are inevitably led
to the conclusion that some process must operate to prevent type
I migration in a standard quiescent disk
 for at least {\em some} protoplanets in order that gas giant
planets can form. 

Although many issues
remain outstanding, 
a number of processes may operate to prevent type I migration.
Being the result of a linear disk response it depends on the 
temperature and density structure of the disk and special
features such as rapid spatial variation of opacity may slow
or stop migration (eg. Menou \& Goodman  2004).
Recent simulations by Nelson \& Papaloizou (2004) and Nelson (2004)
show that low mass planets migrating in magnetised, turbulent accretion disks
undergo stochastic migration rather than monotonic inward
migration. This leads to a distribution of migration rates
for embedded planets, with some undergoing rapid inward migration,
and others perhaps migrating outward or not at all. A well defined direction
of migration is likely to occur when the planet mass is large enough to
dominate over turbulent fluctuations, with simulations indicating that
this is likely to arise for planet masses exceeding  $  \sim 30$ M$_{\oplus}$.
The occurrence of global disk structures such as eccentric $m=1$ modes
are also capable of disrupting both type I (Papaloizou 2002) and
type II (Nelson 2003) migration, and if established within a disk
are likely to be long lived entities. Finally, low mass planets on
eccentric orbits may undergo type I torque reversal (Papaloizou \& Larwood
2000). For an isolated planet the eccentricity is quickly damped,
but a system of mutually interacting planetary cores may be able to 
maintain eccentric orbits and hence reduce or even prevent type I 
migration.

In light of these (and perhaps other) mechanisms for overcoming
type I migration, which may operate in tandem rather than in isolation,
we make the assumption that for masses below $M_{pl} = m_0 
 \simeq 30$ M$_{\oplus}$,
type I migration is essentially ineffective for at least some protoplanets
 below that mass range, such that a population of giant planet can form.
It seems likely that for planet masses larger than this, where the 
disk--planet interaction starts to become non linear, the ability of
the planet to impose itself on the disk will lead to inward migration being
re-established. When it does so figure~\ref{fig13} indicates
it will be at near the maximum rate. The transition from type A to type B
models is near to where $M_{pl}  = m_0.$ Note that $m_0$ is likely to 
depend on location
and parameters in the disk making it uncertain whether the protoplanet
undergoes some rapid inward migration.

Returning to our discussion of the low opacity planet model in
figure~\ref{fig13}, we can see that once the planet mass reaches 
20 M$_{\oplus}$ and moves to a type B track,
 the growth time of the planet decreases dramatically
down to a value that is determined by the rate at which the protostellar
disk can supply mass to the planet.  It is at this stage that we suppose
that the protoplanet undergoes a transition from being an extended
structure in contact with its Roche lobe, and accreting slowly from the 
disk, to a more compact protoplanet with a free surface that accretes
rapidly from the protostellar disk via a circumplanetary accretion disk.
The type B models presented in figures~\ref{fig11} and \ref{fig12} suggest
that these compact models can accrete rapidly from the disk, and we 
specify a growth time of $3 \times 10^3$ years for this stage of growth
in figure~\ref{fig13}, corresponding to the more rapid growth rates 
presented in figure~\ref{fig5}. Such a rate ensures that the planet can grow to
become a Jovian mass gas giant on a time scale shorter than any likely
migration time.

\subsubsection{15 M$_{\oplus}$ core models}
In figure~\ref{fig14} we present a schematic diagram of growth
and migration times for planet models with 15 M$_{\oplus}$ cores.
This figure is very similar to figure~\ref{fig13}.
We have plotted evolutionary sequences for just two planet models
in figure~\ref{fig14}, one with standard opacity (dashed line) and
one with three percent of the standard opacity (dashed-dotted line).
During the earliest stages of accretion, these models have
growth times of $3 \times 10^6$ and $3 \times 10^5$ years, respectively,
which are comfortably
within or below the range of expected disk life-times. In the latter case,
the formation is very rapid, and illustrates the crucial role
played by the opacity. However, the estimated
quiescent disk migration time for an object
with a mass of 15 -- 20 M$_{\oplus}$ is below $10^5$ years, such that even
the lower opacity model is unable to form within the expected type I
migration time. We are again required to assume that type I migration is 
inoperative for some  planets with masses below $ m_0 \simeq 30$ M$_{\oplus}$.

As shown in figure~\ref{fig3}, the growth time for the standard opacity model
presented in figure~\ref{fig14} remains larger than the
corresponding migration time for planet masses up to $\simeq 34$ M$_{\oplus}$,
at which stage rapid gas accretion can ensue.
  Because such  planet models spend longer time at these higher masses
they   may be 
more  susceptible to undergoing
a period of rapid migration either prior to or during the
early stages of rapid gas accretion than are the models with lower core masses.
This may be related to an indication that extrasolar planets
in systems with high metallicity tend to be found at shorter periods
commented on by Santos et al. (2003). However, because of the small
numbers involved, the statistical significance of such a trend is not yet
established.

The result of  a rapid inward migration 
is that the planet will move into the inner regions of
the disk where: ({\it i}) the local reservoir of disk material
is reduced relative to larger radii; ({\it ii}) the disk aspect ratio
$H/R$ decreases making gap formation and a
transition to type II migration  easier (Papaloizou \& Terquem 1999).
The result is likely to be a tendency 
for larger and more dusty cores
to produce a  distribution of planets with a greater
bias toward low mass, short period objects.

\subsection{Final Protoplanet and Circumplanetary Disk  Luminosity }
We have seen that for model sequences of type B, the protoplanet luminosities
are in general smaller and at best comparable
to those expected from the circumplanetary  disk.
In the cases with the most rapid accretion rate the difference is most marked
indicating that the accreting matter in general fails to supply
energy to the protoplanet. This feature causes the radius to attain
and remain near $2 \times 10^{10}$ cm
in most cases. It also means that it is appropriate to regard
protoplanets as undergoing disk accretion and able to  accept all supplied
material  at
reasonable accretion rates once they cease to be enveloped by the disk
as in the type A case. The evolutionary time scales of type A models
are sensitive to the dust opacity, being directly
proportional and  smaller for lower  opacities.

\noindent The reason for the behaviour of type B models
where they fail to expand even at high accretion rates can be
related to some simple properties of barotropic stellar models
that would apply in the completely degenerate limit.
For these $P$ is a specified function of $\rho$ and is related to the internal
energy per unit mass, $U,$  through $P = \rho^2 (d U/d \rho).$
Although the protoplanet models are not of this type, they
are similar enough to make the discussion relevant.

\noindent  The total energy is given by
equation (\ref{toten}). For polytropes of index $n$ and
$U = nP/\rho ,$  it is well known that
(Chandrasekhar 1939)
\begin{equation} E = - {(3-n)GM_{pl}^2\over (5-n)r_s}, \end{equation}
while the mass radius relation is $ r_s \propto M_{pl}^{(n-1)/(n-3)}.$
From this if we consider a small mass  increment $d M_{pl},$ we get
$dE = -(GM_{pl}/ r_s)d M_{pl}.$ But this change represents
the energy lost through disk and boundary layer accretion, leaving
no input for the polytrope explaining why it can remain of low luminosity.
In fact the result is valid for any barotropic model and can
be shown to follow
from the fact that at equilibrium the change of energy
is zero to first order in perturbations (the system can be thought of as being
perturbed from equilibrium at a slightly larger
mass once added material is brought to the
surface). In this way we can understand why models of type B
can remain of low luminosity,
 when compared to the fiducial value
of  $(GM_{pl}/ r_s)d M_{pl}/dt, $  under rapid accretion.

\noindent  The expected circumplanetary 
disk or disk/protoplanet boundary layer luminosity
for a Jovian mass  with radius $2\times 10^{10} cm,$
and final accretion times 
in the range $10^{5-6}y,$ lies  in the range
 $\sim 10^{-(3-4)}  L_{\odot}$ and the characteristic
temperatures are expected to be in the range $1000-2000K.$

\section{Summary \& Conclusion}
 We have presented evolutionary models of giant protoplanets
 forming in protoplanetary disk. First, we have considered planet
 models (type A) consisting of solid rock-ice cores surrounded by gaseous
 envelopes whose surface coincides with the planet Hill sphere,
 and which accrete quasi-statically from the surrounding
 protostellar disk. We have considered models with 5 and 15 M$_{\oplus}$
 cores, and have varied the dust opacity in the envelope.
 For models with a 5 M$_{\oplus}$ core and standard opacity, the
 time required for the planet to undergo rapid gas accretion is
 $\sim 3 \times 10^8$ yr. This is longer than reasonable disk life-times
 ranging between 3 -- 30 Myr. Reductions in the dust opacity by factors
 of 10 and 100 lead to models that require $\sim 3 \times 10^7$ and
 $3 \times 10^6$ yr, respectively, before rapid gas accretion ensues.
 Rapid gas accretion occurs once these planets reach $\sim 18$
 M$_{\oplus}$. A 15 M$_{\oplus}$ core planet model with standard opacity
 takes $\sim 3 \times 10^6$ yr before rapid gas accretion ensues.
 Models with dust opacity reduced by factors of 10 and 100 require
 $\sim 3 \times 10^5$ and $\sim 10^5$ yr before rapid gas accretion
 occurs. This arises once the planet mass exceeds $\sim 35$
 M$_{\oplus}$.

 \noindent We present a second class of planet models (type B) where
 the planet has a free surface, and accretes gas from a circumplanetary
 disk that is fed by the surrounding protostellar disk at a specified
 rate. We find that these models can accrete gas at any reasonable
 rate that may be supplied by the protostellar disk without expansion.

 \noindent We suggest that the earliest stages of giant planet formation
 are described by models of type A. For {\em all} such models,
 the standard type I migration time is shorter than the accretion time
 prior to rapid gas accretion. We suggest that type I migration is
 inoperative for at least some protoplanets with masses below those
 for which disk-planet interactions becomes non-linear (i.e.
 $M_{pl} \sim 30$ M$_{\oplus}$), beyond which planets are more
 likely to undergo rapid inward migration. In such a scenario
 planets with low mass cores and low opacity envelopes will have a
 greater tendency to remain at larger radii up to the point of
 rapid gas accretion. Those with more massive cores will tend to
 undergo more rapid inward migration prior to or during rapid gas
 accretion.

 \noindent At the point of rapid gas accretion, we suppose that planets contract
 within their Hill sphere, and are described by type B models. The
 planets may now accrete at any rate supplied by the protostellar disk,
 and can undergo rapid growth on a time scale shorter than
 the migration time. If planets with low mass cores tend to exist
 at larger radii during this stage, they may make a rapid transition
 to Jovian mass objects, forming gaps and entering a phase of
 slower type II migration. If planets with larger mass cores have
 a tendency to undergo more rapid inward migration, they may
 exist at smaller radii during the rapid gas accretion phase. The disk
 is thinner here -- such that it is easier to form gaps, and the local
 reservoir of gas is smaller. Such objects are more likely to end up with
 sub-Jovian masses.

 \noindent We note that this general picture is likely to be blurred by variations
 in disk parameters and life-times. But we also note that the current
 extrasolar planet data shows a mass-period correlation in line with
 the simple ideas presented here (Zucker \& Mazeh 2002).
 Furthermore there is a hint of a correlation between host star
 metallicity and period such that higher metallicity stars appear to
 host shorter period planets (e.g. Santos et al 2003).   Such a correlation,
 while not statistically significant in the data at present, may turn out to be
 as more data is accumulated and is accordingly a topic for scientific
 consideration (eg. Sozzetti, 2004). We comment that a correlation of this type 
 might
 be expected if planetary core mass and envelope opacity scale with
 the metallicity of the protoplanetary environment.

\label{sec:conclusion}

\begin{acknowledgements} 

 The hydrodynamic simulations performed here
were carried out using the QMUL HPC facility funded
by the SRIF initiative.

\end{acknowledgements}

\end{document}